\documentclass[submission, Phys]{SciPost}

\hypersetup{
    colorlinks,
    linkcolor={red!50!black},
    citecolor={blue!50!black},
    urlcolor={blue!80!black}
}

\usepackage[bitstream-charter]{mathdesign}
\DeclareSymbolFont{usualmathcal}{OMS}{cmsy}{m}{n}
\DeclareSymbolFontAlphabet{\mathcal}{usualmathcal}

\usepackage{amsfonts}
\usepackage[]{amsmath}
\usepackage[]{graphicx}
\usepackage{xcolor}
\usepackage{physics}
\usepackage{bm}
\usepackage{subcaption}
\usepackage[capitalise]{cleveref}
\usepackage{comment}

\graphicspath{{figures/}}

\newcommand\mpb[2]{\{\{ #1, #2 \}\}}

\renewcommand\vec[1]{\overrightarrow{#1}}
\newcommand\cev[1]{\overleftarrow{#1}}

\begin{document}

\begin{center}{\Large \textbf{
  An exact method for bosonizing the Fermi surface in arbitrary dimensions
}}\end{center}

\begin{center}
  Takamori Park\textsuperscript{1$\star$} and
  Leon Balents \textsuperscript{2, 3}
\end{center}

\begin{center}
{\bf 1} Department of Physics,
University of California, Santa Barbara, CA 93106, USA
\\
{\bf 2} Kavli Institute for Theoretical Physics,
University of California, Santa Barbara, CA 93106, USA
\\
{\bf 3} Canadian Institute for Advanced Research, Toronto, Ontario, Canada
\\
${}^\star$ {\small \sf t\_park@ucsb.edu}
\end{center}

\begin{center}
\today
\end{center}


\section*{Abstract}
{\bf
Inspired by the recent work by Delacretaz et. al. \cite{delacretaz2022}, we
rigorously derive an exact and simple method to bosonize a non-interacting
fermionic system with a Fermi surface starting from a microscopic Hamiltonian.
In the long-wavelength limit, we show that the derived bosonized action is
exactly equivalent to the action obtained by Delacretaz et. al. In addition, we
propose diagrammatic rules to evaluate correlation functions using our bosonized
theory and demonstrate these rules by calculating the three- and four-point
density correlation functions. We also consider a general density-density
interaction and show that the simplest approximation in our bosonic theory is
identical to RPA results.
}

\vspace{10pt}
\noindent\rule{\textwidth}{1pt}
\tableofcontents\thispagestyle{fancy}
\noindent\rule{\textwidth}{1pt}
\vspace{10pt}

\section{Introduction}
\label{sec:intro}
Quantum phases of matter with an extensive number of gapless modes form an
interesting class of phases in condensed matter physics that includes the Fermi
liquid phase and the broad class of strongly correlated non-Fermi liquids.
While the former is well understood, non-Fermi liquids are a challenge to study
because they exist in the strong coupling regime which cannot be accessed using
standard perturbative techniques.  Instead, non-perturbative approaches are
necessary to understand how interactions destroy the quasi-particle picture in
non-Fermi liquids. In the past, bosonization of the Fermi surface in $d>1$
attracted attention because of its potential as a non-perturbative method to
study the effects of interactions. This lead to the proposal of several
different higher-dimensional bosonization methods
\cite{luther1979, haldane2005, houghton1993, castroneto1994, castroneto1994a,
houghton1994a, khveshchenko1994, kwon1994, houghton1994, castroneto1995,
frohlich1995, khveshchenko1995, kwon1995, kopietz1995a, kopietz1996}.  However,
these approaches involve complicated constructions (e.g.  separating the Fermi
surface into patches and coarse-graining \cite{haldane2005}). 

More recently, Delacretaz et. al. proposed a new elegant method to bosonize the
Fermi surface that incorporates nonlinear effects to all orders and appears to
correctly capture Fermi liquid physics in the long-wavelength limit
\cite{delacretaz2022}. Guided by the principle that their action should
reproduce the Boltzmann equation, they obtain a bosonic action through a series
of inpsired conjectures.  Due to the heuristic nature of the derivation some
details of their bosonization scheme need clarifying: (a) The bosonized action
is obtained in the long-wavelength limit, but how are corrections to this
incorporated into the theory? (b) Delacretaz et. al.  demonstrated that
tree-level diagrams are sufficient to calculate the two- and three-point density
correlation functions.  Is this generally true for all correlation functions?
What do higher-order loop diagrams contribute, and is there a way to connect
these bosonic diagrams to their corresponding fermionic diagrams?  (c)
Delacretaz et. al. incorporate interaction by writing down the most general form
it can take.  Naively, one would expect a density-density interaction term to
contribute the term $\int V ff$ to the action where $V$ is the potential and $f$
is the distribution function, but is this true?

Inspired by the work by Delacretaz et. al. \cite{delacretaz2022}, in this work
we present an exact
method for the bosonization of a system with a Fermi surface.  In
\cref{sec:derivation}, we outline the derivation of our result.  Our key insight
is to adapt a well known field theory result regarding the effective action that
is commonly found in field theory textbooks \cite{weinberg1995}. Applying this, we are able to
exactly express the generating functional as a restricted path integral over a
bosonic field. In the long-wavelength limit, the action that we obtained reduces
to the action obtained by Delacretaz et. al. We also briefly discuss how
interactions modify the functional integral.  In \cref{sec:diagrams}, we propose
diagrammatic rules to calculate expectation values diagrammatically and
demonstrate their utility by calculating the three- and four-point density
correlation functions. Comparison with the fermion loop result is
straightforward, and we find our result is exactly correct. We also discuss the
effects of interactions and calculate approximate corrections to the two-point
density correlation function.  Finally, in \cref{sec:discussion}, we wrap up our
work with a discussion of our results. We discuss the connection between our
work and the work by Delacretaz et. al. and provide answers to the questions
listed above. In the end we also go over exciting possible future directions 
of our work.

\section{Derivation}
\label{sec:derivation}
In this section, we derive a method to bosonize a non-interacting fermion theory
with a Fermi surface from an algebraic perspective.  We start with a generating
functional and first calculate its Legendre transform.  Then, we will use a
well-known result typically found in field theory textbooks \cite{weinberg1995} to
express the generating functional as a restricted functional integral of its
Legendre transform over a bosonic field.  At the end of this section, we will
also consider how density-density interactions can be incorporated into the
bosonized theory.

As an aside, in our derivation we will deal with functions with two spatial and
one temporal arguments. Initially, we will represent these functions using the
position basis, but this is an arbitrary choice. We could also work in the
momentum basis or a mixed basis (the Wigner representation) \cite{zachos2005,
curtright2014} via an appropriate transformation. It is also possible to handle
these functions in a basis-free manner by noticing, for example, that a function
expressed in the position basis $J(x,x',t)$ can be interpreted as the
position-basis matrix element of a first quantized operator, i.e.  $\exists J$
such that $\mel{x}{J(t)}{x'}=J(x,x',t)$.  As we will see later, this compresses
the notation and simplifies switching between different bases.

\subsection{Generating functional $Z_0[J]$}
We start with the generating functional $Z_0[J]$ that generates density matrix
correlation functions of a non-interacting fermionic system with a Fermi surface
at $T=0$:
\begin{equation}
  Z_0[J]=\expval{U_{[J]}(\infty,-\infty)}{\Omega}.
  \label{eq:genfunct}
\end{equation}
Here, $\ket{\Omega}$ is the many-body ground state representing the Fermi sea
and $U_{[J]}$, which is a functional of the source $J$, is the time-evolution
operator defined in the usual way as
\begin{equation}
  U_{[J]}(t,t')\equiv
  \begin{cases}
    \displaystyle
    \mathbb{T}\exp(-i\int_{t'}^{t}\dd\tau 
    H_0(\tau)-i\int_{t'}^{t}\dd\tau 
    \int_{x,x'}\psi^\dagger(x)J(x,x',\tau)\psi(x')) & t>t'\\[10pt]
    \displaystyle
    \tilde{\mathbb{T}}\exp(-i\int_{t'}^{t}\dd\tau 
    H_0(\tau)-i\int_{t'}^{t}\dd\tau 
    \int_{x,x'}\psi^\dagger(x)J(x,x',\tau)\psi(x')) & t<t'
  \end{cases}.
\end{equation}
$H_0$ is the non-interacting Hamiltonian, and $\mathbb{T},\tilde{\mathbb{T}}$
denote time ordering and anti-time ordering.  We make an important assumption
that the source, $J$, satisfies
\begin{equation}
  U_{[J]}(\infty,-\infty)\ket{\Omega} \propto\ket{\Omega}.
  \label{eq:Jcondition}
\end{equation}
Then, $Z_0[J]$ is simply a complex phase, and if we define the ``free energy''
$F_0[J]$ such that
\begin{equation}
  F_0[J]\equiv i\ln Z_0[J]\quad\qty(\implies Z_0[J]=e^{-iF_0[J]}),
\end{equation}
$F_0[J]$ must be real-valued. Using \cref{eq:Jcondition}, the functional
derivative of $F_0[J]$ is
\begin{align}
  f_{[J]}(x,x',t)\equiv&
  \frac{\delta F_0[J]}{\delta J(x',x,t)}=
  \expval{U_{[J]}(-\infty,t)\psi^\dagger(x')\psi(x)
  U_{[J]}(t,-\infty)}{\Omega}
  \label{eq:fdef}
\end{align}
which is the one-body reduced density matrix initially in the ground state that
evolves under $H_0$ and the source $J$.

Because the Hamiltonian is quadratic in the field operators, the one-body
reduced density matrix which we henceforth simply refer to as the density
matrix, satisfies the von Neumann equation
\begin{equation}
  i\partial_t f_{[J]}(t)=\comm{H_0+J(t)}{f_{[J]}(t)}.
  \label{eq:eom}
\end{equation}
Here and henceforth, $H_0$ refers to the first quantized form of the
non-interacting Hamiltonian.  In addition, because of the condition on $J$,
\cref{eq:Jcondition}, the density matrix (expressed in the momentum basis)
must satisfy the boundary conditions
\begin{equation}
  \lim_{t\rightarrow\pm\infty}f_{[J]}(k,k',t)=\expval{c^\dagger(k')c(k)}{\Omega}
  =f_0(k)(2\pi)^d\delta^d(k-k'),
  \label{eq:fbc}
\end{equation}
where $f_0(k)=\theta(k_F-\abs*{k})$ is the zero-temperature Fermi-Dirac
distribution function.  We assumed a rotationally symmetric Fermi surface for
simplicity. $f_0(k)$ only takes the values 0 or 1, so $f_0^2=f_0$ and $f_0$ must
be a projection operator onto states in the Fermi sea. $f_{[J]}$ evolves
unitarily from $f_0$, so it must also be a projection operator but onto states
in the time-evolved Fermi sea.

\subsection{Legendre transform of $Z_0[J]$}
\label{sec:legendretransform}
Now, let us consider how to calculate the Legendre transform of $Z_0[J]$. As we
will see, the generating functional does not have a well defined Legendre
transform because it is generally not convex \cite{zia2009}. We will resolve this issue by
first restricting $Z_0[J]$ to a domain $\mathcal{F}_s$ in which  it is convex.

Let us begin by first showing that $Z_0[J]$ is not convex.  Consider the map
$J\mapsto f_{[J]}$. This map is not injective because given $f_{[J]}$, if we
define $\chi_{g}\equiv f_{[J]}\chi f_{[J]}+(1-f_{[J]})\chi (1-f_{[J]})\neq0$ for
an aribtrary $\chi$, then $f_{[J]}=f_{[J+\chi_{g}]}$. The proof is
straightforward. First, note that $\comm{\chi_g}{f_{[J]}}=0$ using the property
that $f_{[J]}$ is a projection operator. Then, we can add this term to the
right-hand side of \cref{eq:eom} to get
\begin{equation}
  i\partial_t f_{[J]}(t)=\comm{H_0+J(t)+\chi_g(t)}{f_{[J]}(t)},
  \label{eq:noninjectiveproof}
\end{equation}
but this is the equation of motion for $f_{[J+\chi_g]}$. Since, $f_{[J]}$ and
$f_{[J+\chi_g]}$ both satisfy the same equation of motion and boundary
conditions, it must be true that $f_{[J+\chi_g]}=f_{[J]}$. Therefore, $J\mapsto
f_{[J]}=\frac{\delta F[J]}{\delta J}$ is not an injective map, and $F[J]$ is not
convex.

The transformation $J\rightarrow J+\chi_g$ defines a gauge transformation since
$f_{[J]}$, which is the physical degree of freedom, remains invariant under this
transformation.
If we completely fix this gauge, then $f_{[J]}$ now defines
an injective map. Let $\mathcal{F}_s$ denote the space of sources that satisfies
this gauge and the condition \cref{eq:Jcondition}, and let $\mathcal{F}_d$
denote the set of all density matrices that evolve unitarily in time and satisfy
the boundary conditions in \cref{eq:fbc}.  Then, the map
$\mathcal{F}_s\ni J\mapsto f_{[J]}\in\mathcal{F}_d$
is bijective, and we can
calculate the Legendre transform of $F_0[J]$ defined on $\mathcal{F}_s$.

$f_{[J]}$ defines a bijective map so we can invert it. Given
$f\in\mathcal{F}_d$, we define $J_{[f]}\in\mathcal{F}_s$ such that
$f_{[J]}\eval_{J=J_{[f]}}=f$.  Then, the Legendre transform of 
$F_0[J]$ is 
\begin{align}
  \Gamma_0[f]&\equiv
  \qty(F_0[J]-\int_{x,x',t}\frac{\delta F_0[J]}{\delta J(x',x,t)}J(x',x,t))
  \eval_{J=J_{[f]}}\\
  &=F_0[J_{[f]}]-\Tr[J_{[f]}f]
  \label{eq:Gammadef}
\end{align}
where $\Gamma_0[f]$ is defined for  $f\in\mathcal{F}_d$.  An explicit expression
for $J_{[f]}$ is obtained later in \cref{sec:action}.  The Legendre
transformation is invertible, so there is no loss of information in this
transformation. In fact, the Legendre transformation of $\Gamma_0[f]$ is
$F_0[J]$.

\subsection{Expressing $Z_0[J]$ as a functional integral over a bosonic field}
In this subsection, we derive an expression for the generating functional as a
functional integral over a bosonic field. This is achieved by parametrizing the
space of density matrices $\mathcal{F}_d$ using the Lie algebra of unitary
transformations and adapting a field theory result typically found in
field theory textbooks concerning the effective action \cite{weinberg1995}.

As we discussed in the previous subsection, the Legendre transformation of
$\Gamma_0[f]$ is $F_0[J]$. Using the fact that the leading order term of the
saddle-point approximation is equivalent to the Legendre transform, we can
write \cite{zia2009, weinberg1995}
\begin{equation}
  \int_{\mathcal{F}_d}\mathcal{D}
  f\exp(-i\alpha^{-1}\qty(\Gamma_0[f]+\Tr[fJ]))\rightarrow
  \exp(-i\alpha^{-1}F_0[J]) \qquad \textrm{as}\qquad\alpha\rightarrow0.
  \label{eq:spislegendre}
\end{equation}
It is easy to verify that the saddle-point value of the exponent on the
left-hand side is indeed $F_0[J]$. To evaluate the functional integral in
\cref{eq:spislegendre}, we need to parametrize $\mathcal{F}_d$.  One convenient
choice for parametrization is using the Lie algebra of unitary transformations.
By definition, we know that $f\in\mathcal{F}_d$ evolves unitarily from $f_0$, so
we can always write
\begin{equation}
  f(t)=e^{i\phi(t)}f_0e^{-i\phi(t)}
\end{equation}
where $e^{i\phi(t)}$ is a unitary transformation and $\phi(t)$ is an element of
the Lie algebra of unitary transformations. However, as we discussed in
\cref{sec:legendretransform}, there is a gauge transformation that introduces
redundant descriptions of density matrices. For example, given a unitary
transformation $V(t)$ that commutes with $f_0$, it is clear that
$e^{i\phi(t)}\rightarrow e^{i\phi(t)}V(t)$ is a gauge transformation. We can fix
the gauge by assuming that $\phi\in\mathcal{F}_g$ where $\mathcal{F}_g$ is the
space of all unitary Lie algebra elements that satisfy the conditions
\begin{equation}
  \forall \phi\in\mathcal{F}_g,\, f_0\phi f_0=(1-f_0)\phi (1-f_0)=0
  \quad\textrm{ and }\quad \lim_{t\rightarrow\pm\infty}\phi(t)=0.
  \label{eq:phicondition}
\end{equation}
The first condition implies $\phi$ only couples momentum states inside and
outside of the Fermi sea. There are other ways to fix the gauge, but this is the
most convenient. The second condition in \cref{eq:phicondition} imposes the
boundary condition for $f$, \cref{eq:fbc}.

Using this parametrization, $f=f[\phi]$ is now a functional of $\phi$, and the
functional integral in \cref{eq:spislegendre} in the limit of small $\alpha$ is
\begin{equation}
  I[J,\alpha]\equiv
  \int_{\mathcal{F}_g}\mathcal{D}\phi\exp(-i\alpha^{-1}\qty(\Gamma_0[f[\phi]]
  +\Tr[f[\phi]J])).
\end{equation}
The Jacobian $\mathcal{J}[\phi]= \det[\frac{\delta f[\phi]}{\delta \phi}]$ that
comes from the change of variables is a subleading contribution to the action as
$\alpha\rightarrow0$, so it can be ignored.  Technically, the unitary group is
compact, but extending the integral over $\phi$ to all values only changes the
functional integral up to an overall factor. Therefore, $I[J,\alpha]$ can now be
treated as a Gaussian integral with perturbations, and it can be evaluated using
Feynman diagrams.

Let us consider what type of diagrams we need to consider as
$\alpha\rightarrow0$. It is known that $\alpha$ is a loop expansion parameter:
Each vertex contributes a factor of $\alpha^{-1}$ and each propagator
contributes a factor of $\alpha$, so Feynman diagrams with $L$ loops are
proportional to $\alpha^{L-1}$.  Since $I[J,\alpha]\rightarrow
\exp(-i\alpha^{-1}F_0[J])$ as $\alpha\rightarrow0$, $-iF_0[J]$ must correspond
to connected diagrams with the weight $\alpha^{-1}$ which are tree diagrams
(L=0). Therefore,
\begin{align}
  F_{0}[J]=&i\underset{\substack{\textrm{connected}\\ \textrm{tree}}}
  {\int_{\mathcal{F}_g}}
  \mathcal{D}\phi e^{-i(\Gamma_0[f[\phi]]-\Gamma_0[f_0])
  -i\Tr[(f[\phi]-f_0)J]}+\Tr[f_0 J]
\end{align}
up to an additive constant, and the generating functional itself is the sum of
all tree diagrams (that don't need to be connected),
\begin{equation}
  Z_0[J]=\underset{\textrm{tree}}{\int_{\mathcal{F}_g}}
  \mathcal{D}\phi e^{-i\Gamma_0[f[\phi]]-i\Tr[f[\phi]J]}.
\end{equation}
We refer to these functional integrals that are limited to certain diagrams as
restricted functional integrals.  As we have already mentioned, this is an
adaptation of a well-known field theory result on effective actions found in
field theory textbooks \cite{weinberg1995}.

\subsection{Bosonized action $S[\phi]$}
\label{sec:action}
Here, we define the bosonized action $S[\phi]$ and find an explicit expression
for it. We also consider how interactions modify the action.

The definition for the action is given by
\begin{equation}
  S[\phi]=-\Gamma_0[f[\phi]]+\Gamma_0[f_0].
\end{equation}
In order to find an explicit expression for $\Gamma_0[f[\phi]]$, we first
identify the correspondence
\begin{equation}
  U_{[J_{[f]}]}(t,-\infty)=e^{i\phi(t)}.
  \label{eq:Uphicorrespondence}
\end{equation}
There are other ways to relate $U$ and $\phi$ because of the gauge structure,
but this is most convenient. Under this correspondence, it is clear that
$Z_0[J_{[f]}]=\expval{U_{[J_{[f]}]}(\infty,-\infty)}{\Omega}=1$ and
$F_0[J_{[f]}]=0$ because of the second condition in \cref{eq:phicondition}.
Therefore, the first term
in \cref{eq:Gammadef} vanishes. Next, taking the
time derivative of \cref{eq:Uphicorrespondence} gives us $i\partial_t
e^{i\phi(t)}=(H_0+J_{f[t]}(t))e^{i\phi(t)}$, so
\begin{equation}
  J_{[f]}=i(\partial_t e^{i\phi(t)})e^{-i\phi(t)}-H_0.
\end{equation}
Therefore, the action must be
\begin{equation}
  S[\phi]=
  \Tr[J_{[f]}f[\phi]]+\Tr[H_0f_0]
  =\Tr[f_0e^{-i\phi}(i\partial_t-H_0)e^{i\phi}]+\Tr[f_0H_0].
  \label{eq:action}
\end{equation}
The free energy and generating functional is given by
\begin{align}
  F_0[J]=&i\underset{\textrm{connected tree}}{\int_{\mathcal{F}_g}}
  \mathcal{D}\phi e^{iS[\phi]-i\Tr[(f[\phi]-f_0)J]}+\Tr[f_0 J]\\
  Z_0[J]=&\underset{\textrm{tree}}{\int_{\mathcal{F}_g}}
  \mathcal{D}\phi e^{iS[\phi]-i\Tr[f[\phi]J]}.
\end{align}

Up until now, we have exclusively considered non-interacting systems, but
incorporating interactions is straightforward. Consider a general two-body
interaction
\begin{equation}
  H_{\textrm{int}}=\frac{1}{2}\int_{x,y}V(x-y)
  \psi^\dagger(x)\psi^\dagger(y)\psi(y)\psi(x).
\end{equation}
The generating functional of the interacting system $Z[J]$ can be expressed
using its non-interacting counter part according to
\begin{equation}
  Z[J]=\exp(\frac{i}{2}\int_{x,y,t}V(x-y)
  \frac{\delta}{\delta J(x,x,t)}\frac{\delta}{\delta J(y,y,t)})Z_0[J].
\end{equation}
This adds an additional term to the action $S[\phi]\rightarrow
S[\phi]+S_\textrm{int}[\phi]$ where
\begin{align}
  S_{\textrm{int}}[\phi]=&-\frac{1}{2}\int_{x,y,t}V(x-y)
  f[\phi](x,x,t)f[\phi](y,y,t)\\
  =&-\frac{1}{2}\int_{q,k,k',t}\tilde{V}_qf[\phi](k+q,k,t)f[\phi](k'-q,k',t).
  \label{eq:interaction}
\end{align}

The action that we obtained (\cref{eq:action}) looks similar to the action
derived by Delacretaz et. al. \cite{delacretaz2022}. In
\cref{sec:discussion} and \cref{sec:correspondence} we discuss the link between
the two actions in detail.

\section{Diagrammatic calculation of correlation functions}
\label{sec:diagrams}
In this section, we propose a set of diagrammatic rules that simplifies the task
of calculating correlation functions from the bosonic theory. We demonstrate
their use by calculating the three- and four-point density correlation
functions.  Since the bosonic theory is exact, we can directly compare results
with the equivalent fermionic theory. We find that the diagrams obtained from
the bosonic theory are partitions of the corresponding diagrams obtained from
the fermionic theory. In other words, the process of bosonization separates each
fermion diagrams into several bosonic diagrams. In the long wavelength limit,
fermion 1-loop diagrams exhibit partial cancellation and extracting the correct
leading order behavior can be tricky \cite{neumayr1998, holder2015}. However, we
find that the diagrams calculated from the bosonic theory are well-behaved, and
calculating the correct long wavelength behavior becomes straightforward.

\subsection{Expanding the action and density matrix in powers of $\phi$}
We first begin by expanding the action given in \cref{eq:action} in powers of
$\phi$. We find
\begin{equation}
  S[\phi]=\sum_{n=1}^{\infty}\frac{(-i)^n}{n!}
  \Tr[f_0\qty(-i\textrm{ad}_\phi^{n-1}\partial_t\phi-\textrm{ad}_\phi^nH_0)],
\end{equation}
where $\textrm{ad}_\phi(\cdot)\equiv\comm{\phi}{\cdot}$ is the adjoint action of
the Lie algebra. The action is most conveniently expressed in the momentum basis
since $H_0$ and $f_0$ are diagonal in this basis. The symmetrized action in the
momentum basis is
\begin{align}
  S[\phi]=\sum_{\substack{n=2\\n:\text{even}}}^{\infty}S_{(n)}[\phi],
\end{align}
where 
\begin{align}
  S_{(n)}[\phi]\equiv \frac{i^n2^{n-2}}{n!}
  \int_{\substack{p_1,\cdots,p_n\\ \omega_1,\cdots,\omega_n}}&
  \qty[\frac{-\omega_1+\omega_2-\cdots+\omega_n}{n}
  -2\frac{-\varepsilon_{p_1}+\varepsilon_{p_2}-\cdots+\varepsilon_{p_n}}{n}]
   F^{(n)}(p_1,\cdots,p_n)\notag\\
   &\qquad \times\phi(p_1,p_2,\omega_1)\cdots\phi(p_n,p_1,\omega_n)
  2\pi\delta(\omega_1+\cdots+\omega_n).
\end{align}
$\varepsilon_{p}=\expval{H_0}{p}$ is the dispersion which for simplicity we
have assumed is $\varepsilon_{p}=\frac{p^2}{2m}-E_F$. The factor $F^{(n)}$ is a
function that only takes the values $-1,0,1$ and is defined as
\begin{align}
  F^{(n)}(p_1,\cdots,p_{n})\equiv
  \begin{cases}
    \begin{aligned}
    &f_0(p_1)[1-f_0(p_2)]f_0(p_3)\cdots f_0(p_n)\\
    &\qquad-[1-f_0(p_0)]f_0(p_1)[1-f_0(p_2)]\cdots[1-f_0(p_n)]
    \end{aligned}
    & n\text{: even}\\[10pt]
    \begin{aligned}
    &[1-f_0(p_1)]f_0(p_2)[1-f_0(p_3)]\cdots f_0(p_n)\\
    &\qquad-f_0(p_1)[1-f_0(p_2)]f_0(p_3)\cdots[1-f_0(p_n)]
    \end{aligned}
    & n\text{: odd}
  \end{cases}.
  \label{eq:Fdef}
\end{align}
Adjacent momenta, $p_i,p_{i+1}$, in the argument cannot both be in or out of the
Fermi sea. This is a consequence of the first condition in
\cref{eq:phicondition}. For example, if $f_0(p_{i})=f_0(p_{i+1})$, then
$F^{(n)}(p_1,\cdots,p_n)=0$. In addition, as a result of this condition, odd $n$
terms in the action must vanish because they are proportional to
$f_0(p_1)(1-f_0(p_1))$ which vanishes because $f_0$ is a projection operator.

The quadratic contribution to the action is
\begin{equation}
  S_0[\phi]\equiv S_{(2)}[\phi]=
  \int_{k,k'}\frac{1}{2}\qty[\omega-\varepsilon_k+\varepsilon_k']
  \abs{\phi(k,k',\omega)}^2F^{(2)}(k,k').
  \label{eq:quadraticaction}
\end{equation}
This allows us to define the propagator evaluated with respect to $S_0$
\begin{align}
  \expval{\phi(k,k',\omega)\phi(p',p,\omega')}_0=&D_0(k,k',\omega)
  (2\pi)^{2d+1}\delta^d(p-k)\delta^d(p'-k')\delta(\omega+\omega').
\end{align}
where
\begin{equation}
  D_0(k,k',\omega)=D_0(k',k,-\omega)\equiv
  -i\frac{f_0(k)-f_0(k')}{\omega-\varepsilon_{k}+\varepsilon_{k'}}
\end{equation}
Here, we used the fact that, $F^{(2)}(k,k')=f_0(k')-f_0(k)$.
Note, $\expval{\cdots}_0$ indicates evaluation with respect to $S_0$.

Next, starting from our choice of the definition
$f(t)=e^{i\phi(t)}f_0e^{-i\phi(t)}$, we also expand the density matrix in powers
of $\phi$ and find
\begin{equation}
  f[\phi](t)=\sum_{n=0}^\infty\frac{i^n}{n!}\textrm{ad}_{\phi(t)}^nf_0.
\end{equation}
In the momentum basis, the density matrix is expressed as
\begin{equation}
  f[\phi](k,k',t)=\sum_{n=0}^\infty f_{(n)}[\phi](k,k',t)
\end{equation}
where
\begin{align}
  f_{(n)}(k,k',t)\equiv\frac{i^n2^{n-1}}{n!}
  \int_{k_1,\cdots,k_{n-1}}
  \phi(k,k_1,t)\phi(k_1,k_2,t)\cdots
  \phi(k_{n-1},k',t)F^{(n+1)}(k,k_1,\cdots,k_{n-1},k').
\end{align}

Lastly, let us go over how to evaluate correlation functions.  Consider a general
correlation function of the form
\begin{equation}
  \expval{f(k_1,k_1',\omega_1)\cdots f(k_n,k_n',\omega_n)}.
\end{equation}
The general procedure to evaluate such a correlation function is to expand the
action and each $f$ in powers of $\phi$ and keep only the tree diagrams. 
Explicitly,
\begin{equation}
  \begin{split}
  &\expval{f(k_1,k_1',\omega_1)\cdots f(k_N,k_N',\omega_N)}\\
  &\qquad\qquad=\sum_{n_1=0}^\infty\cdots\sum_{n_N=0}^\infty
  \sum_{\substack{m=4\\m:\textrm{even}}}^\infty
  \expval{f_{(n_1)}(k_1,k_1',\omega_1)\cdots f_{(n_N)}(k_N,k_N',\omega_N)i
  S_{(m)}}_{0,c,\textrm{tree}}
  \end{split}
\end{equation}
Even though, the expansion contains an infinite number of terms, since the
expectation value is restricted to tree diagrams, in reality we only need to
consider a finite number of terms.

\subsection{Diagrammatic rules}
\begin{figure}[h]
  \centering
  \hfill
  \subcaptionbox{\label{fig:phi}}
  {
  \def\svgwidth{0.3\textwidth}
\begingroup%
  \makeatletter%
  \providecommand\color[2][]{%
    \errmessage{(Inkscape) Color is used for the text in Inkscape, but the package 'color.sty' is not loaded}%
    \renewcommand\color[2][]{}%
  }%
  \providecommand\transparent[1]{%
    \errmessage{(Inkscape) Transparency is used (non-zero) for the text in Inkscape, but the package 'transparent.sty' is not loaded}%
    \renewcommand\transparent[1]{}%
  }%
  \providecommand\rotatebox[2]{#2}%
  \newcommand*\fsize{\dimexpr\f@size pt\relax}%
  \newcommand*\lineheight[1]{\fontsize{\fsize}{#1\fsize}\selectfont}%
  \ifx\svgwidth\undefined%
    \setlength{\unitlength}{198.42519685bp}%
    \ifx\svgscale\undefined%
      \relax%
    \else%
      \setlength{\unitlength}{\unitlength * \real{\svgscale}}%
    \fi%
  \else%
    \setlength{\unitlength}{\svgwidth}%
  \fi%
  \global\let\svgwidth\undefined%
  \global\let\svgscale\undefined%
  \makeatother%
  \begin{picture}(1,0.71428571)%
    \lineheight{1}%
    \setlength\tabcolsep{0pt}%
    \put(0,0){\includegraphics[width=\unitlength,page=1]{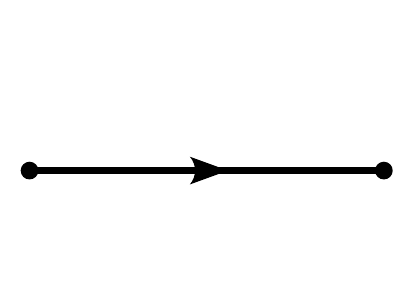}}%
    \put(0.89682433,0.22362591){\color[rgb]{0,0,0}\makebox(0,0)[lt]{\lineheight{1.25}\smash{\begin{tabular}[t]{l}$k$\end{tabular}}}}%
    \put(0.02006738,0.22362591){\color[rgb]{0,0,0}\makebox(0,0)[lt]{\lineheight{1.25}\smash{\begin{tabular}[t]{l}$k'$\end{tabular}}}}%
    \put(0,0){\includegraphics[width=\unitlength,page=2]{phi.pdf}}%
    \put(0.54421257,0.43244547){\color[rgb]{0,0,0}\makebox(0,0)[lt]{\lineheight{1.25}\smash{\begin{tabular}[t]{l}$k-k',\omega$\end{tabular}}}}%
    \put(0.344828,0.16968916){\color[rgb]{0,0,0}\makebox(0,0)[lt]{\lineheight{1.25}\smash{\begin{tabular}[t]{l}$\phi(k,k',\omega)$\end{tabular}}}}%
  \end{picture}%
\endgroup%

  }
  \hfill
  \subcaptionbox{\label{fig:propagator}}
  {
  \def\svgwidth{0.3\textwidth}
\begingroup%
  \makeatletter%
  \providecommand\color[2][]{%
    \errmessage{(Inkscape) Color is used for the text in Inkscape, but the package 'color.sty' is not loaded}%
    \renewcommand\color[2][]{}%
  }%
  \providecommand\transparent[1]{%
    \errmessage{(Inkscape) Transparency is used (non-zero) for the text in Inkscape, but the package 'transparent.sty' is not loaded}%
    \renewcommand\transparent[1]{}%
  }%
  \providecommand\rotatebox[2]{#2}%
  \newcommand*\fsize{\dimexpr\f@size pt\relax}%
  \newcommand*\lineheight[1]{\fontsize{\fsize}{#1\fsize}\selectfont}%
  \ifx\svgwidth\undefined%
    \setlength{\unitlength}{198.42519685bp}%
    \ifx\svgscale\undefined%
      \relax%
    \else%
      \setlength{\unitlength}{\unitlength * \real{\svgscale}}%
    \fi%
  \else%
    \setlength{\unitlength}{\svgwidth}%
  \fi%
  \global\let\svgwidth\undefined%
  \global\let\svgscale\undefined%
  \makeatother%
  \begin{picture}(1,0.71428571)%
    \lineheight{1}%
    \setlength\tabcolsep{0pt}%
    \put(0,0){\includegraphics[width=\unitlength,page=1]{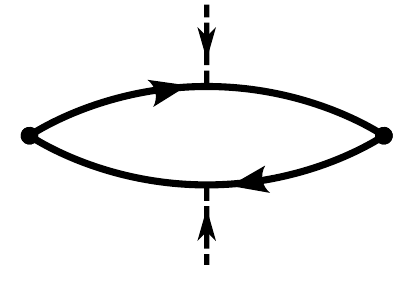}}%
    \put(-0.05090894,0.37116052){\color[rgb]{0,0,0}\makebox(0,0)[lt]{\lineheight{1.25}\smash{\begin{tabular}[t]{l}$k'$\end{tabular}}}}%
    \put(0.95003996,0.37532102){\color[rgb]{0,0,0}\makebox(0,0)[lt]{\lineheight{1.25}\smash{\begin{tabular}[t]{l}$k$\end{tabular}}}}%
    \put(0.53916368,0.55877004){\color[rgb]{0,0,0}\makebox(0,0)[lt]{\lineheight{1.25}\smash{\begin{tabular}[t]{l}$k-k', \omega$\end{tabular}}}}%
    \put(0.52775795,0.16561987){\color[rgb]{0,0,0}\makebox(0,0)[lt]{\lineheight{1.25}\smash{\begin{tabular}[t]{l}$k'-k, -\omega$\end{tabular}}}}%
    \put(0.32732722,0.01884715){\color[rgb]{0,0,0}\makebox(0,0)[lt]{\lineheight{1.25}\smash{\begin{tabular}[t]{l}$D_0(k,k',\omega)$\end{tabular}}}}%
  \end{picture}%
\endgroup%

  }
  \hfill\null
  \caption{Feynman diagram elements for (a) the field $\phi(k,k',\omega)$ and
  (b) the propagator $D_0(k,k',\omega)$.}
  \label{fig:diagramelements}
\end{figure}

Correlation functions can be calculated using the bosonized theory without the
use of diagrams, but they often involve lengthy and tedious algebra resulting
from integrating over multiple delta functions. This is because the field $\phi$
has two spatial indices rather than one which doubles the number of spatial
delta functions. This can be avoided by using Feynman diagrams. Below, we
establish Feynman diagram rules for the bosonized theory:
\begin{itemize}
  \item Points represent momenta.
  \item The field $\phi(k,k',\omega)$ is represented by a line with an arrow
    that starts at $k'$ and ends at $k$. A dashed line with an arrow is used to
    keep track of the frequency and momentum exchange. (\cref{fig:phi})
  \item The propagator $D_0(k,k',\omega)$ is represented by a loop connecting the
    momenta $k$ and $k'$. (\cref{fig:propagator})
\end{itemize}

\subsection{Three-point density correlation function}
\label{sec:threepoint}
\begin{figure}[h]
  \centering
  \hfill
  \subcaptionbox{\label{fig:schm3}}{
    \def\svgwidth{0.18\textwidth}
\begingroup%
  \makeatletter%
  \providecommand\color[2][]{%
    \errmessage{(Inkscape) Color is used for the text in Inkscape, but the package 'color.sty' is not loaded}%
    \renewcommand\color[2][]{}%
  }%
  \providecommand\transparent[1]{%
    \errmessage{(Inkscape) Transparency is used (non-zero) for the text in Inkscape, but the package 'transparent.sty' is not loaded}%
    \renewcommand\transparent[1]{}%
  }%
  \providecommand\rotatebox[2]{#2}%
  \newcommand*\fsize{\dimexpr\f@size pt\relax}%
  \newcommand*\lineheight[1]{\fontsize{\fsize}{#1\fsize}\selectfont}%
  \ifx\svgwidth\undefined%
    \setlength{\unitlength}{198.42519685bp}%
    \ifx\svgscale\undefined%
      \relax%
    \else%
      \setlength{\unitlength}{\unitlength * \real{\svgscale}}%
    \fi%
  \else%
    \setlength{\unitlength}{\svgwidth}%
  \fi%
  \global\let\svgwidth\undefined%
  \global\let\svgscale\undefined%
  \makeatother%
  \begin{picture}(1,1)%
    \lineheight{1}%
    \setlength\tabcolsep{0pt}%
    \put(0,0){\includegraphics[width=\unitlength,page=1]{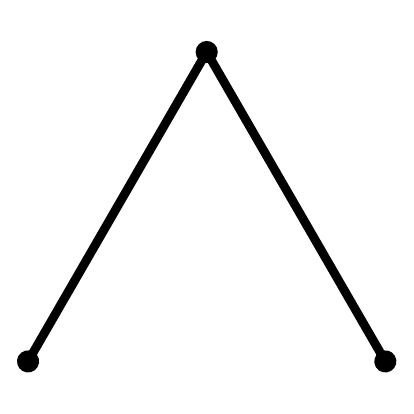}}%
    \put(-0.020576,0.04774762){\color[rgb]{0,0,0}\makebox(0,0)[lt]{\lineheight{1.25}\smash{\begin{tabular}[t]{l}$\rho_{(1)}$\end{tabular}}}}%
    \put(0.41959526,0.93603194){\color[rgb]{0,0,0}\makebox(0,0)[lt]{\lineheight{1.25}\smash{\begin{tabular}[t]{l}$\rho_{(2)}$\end{tabular}}}}%
    \put(0.85363056,0.04774762){\color[rgb]{0,0,0}\makebox(0,0)[lt]{\lineheight{1.25}\smash{\begin{tabular}[t]{l}$\rho_{(1)}$\end{tabular}}}}%
  \end{picture}%
\endgroup%
 
  }
  \hfill
  \subcaptionbox{\label{fig:schm41}}{
    \def\svgwidth{0.18\textwidth}
\begingroup%
  \makeatletter%
  \providecommand\color[2][]{%
    \errmessage{(Inkscape) Color is used for the text in Inkscape, but the package 'color.sty' is not loaded}%
    \renewcommand\color[2][]{}%
  }%
  \providecommand\transparent[1]{%
    \errmessage{(Inkscape) Transparency is used (non-zero) for the text in Inkscape, but the package 'transparent.sty' is not loaded}%
    \renewcommand\transparent[1]{}%
  }%
  \providecommand\rotatebox[2]{#2}%
  \newcommand*\fsize{\dimexpr\f@size pt\relax}%
  \newcommand*\lineheight[1]{\fontsize{\fsize}{#1\fsize}\selectfont}%
  \ifx\svgwidth\undefined%
    \setlength{\unitlength}{198.42519685bp}%
    \ifx\svgscale\undefined%
      \relax%
    \else%
      \setlength{\unitlength}{\unitlength * \real{\svgscale}}%
    \fi%
  \else%
    \setlength{\unitlength}{\svgwidth}%
  \fi%
  \global\let\svgwidth\undefined%
  \global\let\svgscale\undefined%
  \makeatother%
  \begin{picture}(1,1)%
    \lineheight{1}%
    \setlength\tabcolsep{0pt}%
    \put(0,0){\includegraphics[width=\unitlength,page=1]{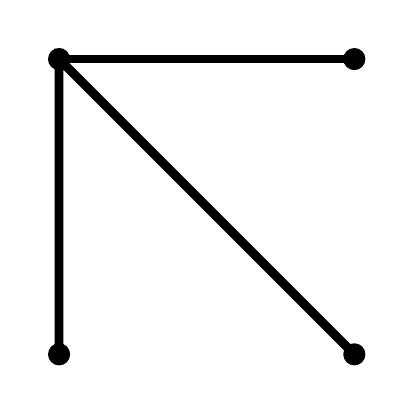}}%
    \put(0.73824873,0.92285463){\color[rgb]{0,0,0}\makebox(0,0)[lt]{\lineheight{1.25}\smash{\begin{tabular}[t]{l}$\rho_{(1)}$\end{tabular}}}}%
    \put(0.72728146,0.05150396){\color[rgb]{0,0,0}\makebox(0,0)[lt]{\lineheight{1.25}\smash{\begin{tabular}[t]{l}$\rho_{(1)}$\end{tabular}}}}%
    \put(0.05757176,0.05150396){\color[rgb]{0,0,0}\makebox(0,0)[lt]{\lineheight{1.25}\smash{\begin{tabular}[t]{l}$\rho_{(1)}$\end{tabular}}}}%
    \put(0.05757176,0.92285463){\color[rgb]{0,0,0}\makebox(0,0)[lt]{\lineheight{1.25}\smash{\begin{tabular}[t]{l}$\rho_{(3)}$\end{tabular}}}}%
  \end{picture}%
\endgroup%
 
  }
  \hfill
  \subcaptionbox{\label{fig:schm42}}{
    \def\svgwidth{0.18\textwidth}
\begingroup%
  \makeatletter%
  \providecommand\color[2][]{%
    \errmessage{(Inkscape) Color is used for the text in Inkscape, but the package 'color.sty' is not loaded}%
    \renewcommand\color[2][]{}%
  }%
  \providecommand\transparent[1]{%
    \errmessage{(Inkscape) Transparency is used (non-zero) for the text in Inkscape, but the package 'transparent.sty' is not loaded}%
    \renewcommand\transparent[1]{}%
  }%
  \providecommand\rotatebox[2]{#2}%
  \newcommand*\fsize{\dimexpr\f@size pt\relax}%
  \newcommand*\lineheight[1]{\fontsize{\fsize}{#1\fsize}\selectfont}%
  \ifx\svgwidth\undefined%
    \setlength{\unitlength}{198.42519685bp}%
    \ifx\svgscale\undefined%
      \relax%
    \else%
      \setlength{\unitlength}{\unitlength * \real{\svgscale}}%
    \fi%
  \else%
    \setlength{\unitlength}{\svgwidth}%
  \fi%
  \global\let\svgwidth\undefined%
  \global\let\svgscale\undefined%
  \makeatother%
  \begin{picture}(1,1)%
    \lineheight{1}%
    \setlength\tabcolsep{0pt}%
    \put(0,0){\includegraphics[width=\unitlength,page=1]{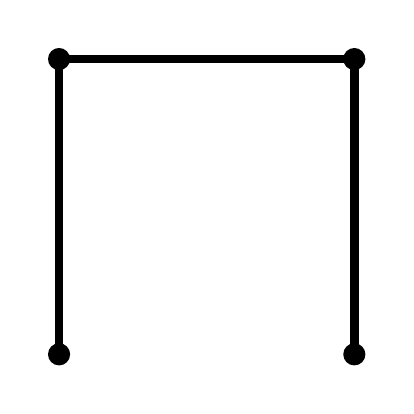}}%
    \put(0.0414362,0.92224631){\color[rgb]{0,0,0}\makebox(0,0)[lt]{\lineheight{1.25}\smash{\begin{tabular}[t]{l}$\rho_{(2)}$\end{tabular}}}}%
    \put(0.75123137,0.92224631){\color[rgb]{0,0,0}\makebox(0,0)[lt]{\lineheight{1.25}\smash{\begin{tabular}[t]{l}$\rho_{(2)}$\end{tabular}}}}%
    \put(0.75123137,0.06328768){\color[rgb]{0,0,0}\makebox(0,0)[lt]{\lineheight{1.25}\smash{\begin{tabular}[t]{l}$\rho_{(1)}$\end{tabular}}}}%
    \put(0.0414362,0.0521034){\color[rgb]{0,0,0}\makebox(0,0)[lt]{\lineheight{1.25}\smash{\begin{tabular}[t]{l}$\rho_{(1)}$\end{tabular}}}}%
  \end{picture}%
\endgroup%
 
  }
  \hfill
  \subcaptionbox{\label{fig:schm43}}{
    \def\svgwidth{0.18\textwidth}
\begingroup%
  \makeatletter%
  \providecommand\color[2][]{%
    \errmessage{(Inkscape) Color is used for the text in Inkscape, but the package 'color.sty' is not loaded}%
    \renewcommand\color[2][]{}%
  }%
  \providecommand\transparent[1]{%
    \errmessage{(Inkscape) Transparency is used (non-zero) for the text in Inkscape, but the package 'transparent.sty' is not loaded}%
    \renewcommand\transparent[1]{}%
  }%
  \providecommand\rotatebox[2]{#2}%
  \newcommand*\fsize{\dimexpr\f@size pt\relax}%
  \newcommand*\lineheight[1]{\fontsize{\fsize}{#1\fsize}\selectfont}%
  \ifx\svgwidth\undefined%
    \setlength{\unitlength}{198.42519685bp}%
    \ifx\svgscale\undefined%
      \relax%
    \else%
      \setlength{\unitlength}{\unitlength * \real{\svgscale}}%
    \fi%
  \else%
    \setlength{\unitlength}{\svgwidth}%
  \fi%
  \global\let\svgwidth\undefined%
  \global\let\svgscale\undefined%
  \makeatother%
  \begin{picture}(1,1)%
    \lineheight{1}%
    \setlength\tabcolsep{0pt}%
    \put(0,0){\includegraphics[width=\unitlength,page=1]{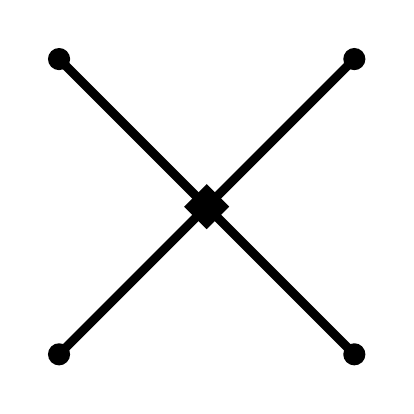}}%
    \put(0.04537611,0.92444951){\color[rgb]{0,0,0}\makebox(0,0)[lt]{\lineheight{1.25}\smash{\begin{tabular}[t]{l}$\rho_{(1)}$\end{tabular}}}}%
    \put(0.04537611,0.05239379){\color[rgb]{0,0,0}\makebox(0,0)[lt]{\lineheight{1.25}\smash{\begin{tabular}[t]{l}$\rho_{(1)}$\end{tabular}}}}%
    \put(0.72312536,0.05239379){\color[rgb]{0,0,0}\makebox(0,0)[lt]{\lineheight{1.25}\smash{\begin{tabular}[t]{l}$\rho_{(1)}$\end{tabular}}}}%
    \put(0.72312536,0.92444951){\color[rgb]{0,0,0}\makebox(0,0)[lt]{\lineheight{1.25}\smash{\begin{tabular}[t]{l}$\rho_{(1)}$\end{tabular}}}}%
    \put(0.58962855,0.48842165){\color[rgb]{0,0,0}\makebox(0,0)[lt]{\lineheight{1.25}\smash{\begin{tabular}[t]{l}$S_{(4)}$\end{tabular}}}}%
  \end{picture}%
\endgroup%
 
  }
  \hfill\null
  \caption{Schematic tree-level diagrams for the three- and four-point density
  correlation functions.  (a) corresponds to the three-point correlation
  function and (b--e) correspond to the four-point correlation function.}
  \label{fig:schematics}
\end{figure}
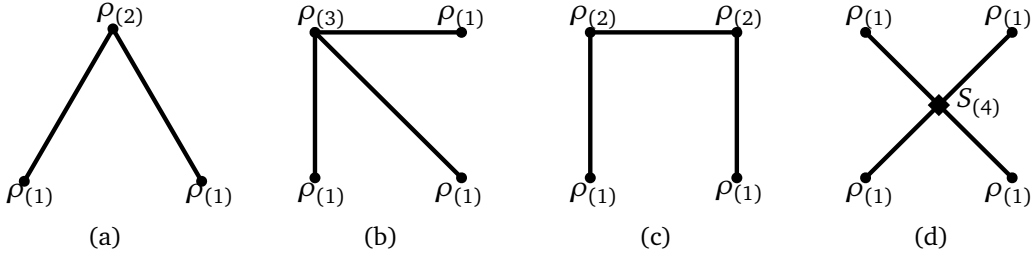

\begin{figure}[h]
  \centering
  \hfill
  \subcaptionbox{\label{fig:c3a}}{
    \def\svgwidth{0.35\textwidth}
\begingroup%
  \makeatletter%
  \providecommand\color[2][]{%
    \errmessage{(Inkscape) Color is used for the text in Inkscape, but the package 'color.sty' is not loaded}%
    \renewcommand\color[2][]{}%
  }%
  \providecommand\transparent[1]{%
    \errmessage{(Inkscape) Transparency is used (non-zero) for the text in Inkscape, but the package 'transparent.sty' is not loaded}%
    \renewcommand\transparent[1]{}%
  }%
  \providecommand\rotatebox[2]{#2}%
  \newcommand*\fsize{\dimexpr\f@size pt\relax}%
  \newcommand*\lineheight[1]{\fontsize{\fsize}{#1\fsize}\selectfont}%
  \ifx\svgwidth\undefined%
    \setlength{\unitlength}{340.15748031bp}%
    \ifx\svgscale\undefined%
      \relax%
    \else%
      \setlength{\unitlength}{\unitlength * \real{\svgscale}}%
    \fi%
  \else%
    \setlength{\unitlength}{\svgwidth}%
  \fi%
  \global\let\svgwidth\undefined%
  \global\let\svgscale\undefined%
  \makeatother%
  \begin{picture}(1,0.66666667)%
    \lineheight{1}%
    \setlength\tabcolsep{0pt}%
    \put(0,0){\includegraphics[width=\unitlength,page=1]{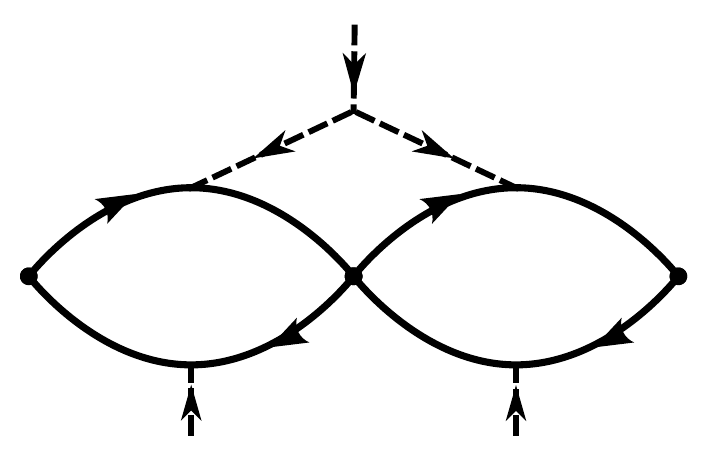}}%
    \put(0.54367486,0.57359851){\color[rgb]{0,0,0}\makebox(0,0)[lt]{\lineheight{1.25}\smash{\begin{tabular}[t]{l}$q_1,\omega_1$\end{tabular}}}}%
    \put(0.30109773,0.08520181){\color[rgb]{0,0,0}\makebox(0,0)[lt]{\lineheight{1.25}\smash{\begin{tabular}[t]{l}$q_3,\omega_3$\end{tabular}}}}%
    \put(0.75943685,0.08520181){\color[rgb]{0,0,0}\makebox(0,0)[lt]{\lineheight{1.25}\smash{\begin{tabular}[t]{l}$q_2,\omega_2$\end{tabular}}}}%
    \put(-0.02645833,0.26939549){\color[rgb]{0,0,0}\makebox(0,0)[lt]{\lineheight{1.25}\smash{\begin{tabular}[t]{l}$k$\end{tabular}}}}%
    \put(0.98998261,0.2696107){\color[rgb]{0,0,0}\makebox(0,0)[lt]{\lineheight{1.25}\smash{\begin{tabular}[t]{l}$k+q_1$\end{tabular}}}}%
    \put(0.54606869,0.2696107){\color[rgb]{0,0,0}\makebox(0,0)[lt]{\lineheight{1.25}\smash{\begin{tabular}[t]{l}$k+q_1+q_2$\end{tabular}}}}%
  \end{picture}%
\endgroup%
 
  }
  \hfill
  \subcaptionbox{\label{fig:c3b}}{
    \def\svgwidth{0.35\textwidth}
\begingroup%
  \makeatletter%
  \providecommand\color[2][]{%
    \errmessage{(Inkscape) Color is used for the text in Inkscape, but the package 'color.sty' is not loaded}%
    \renewcommand\color[2][]{}%
  }%
  \providecommand\transparent[1]{%
    \errmessage{(Inkscape) Transparency is used (non-zero) for the text in Inkscape, but the package 'transparent.sty' is not loaded}%
    \renewcommand\transparent[1]{}%
  }%
  \providecommand\rotatebox[2]{#2}%
  \newcommand*\fsize{\dimexpr\f@size pt\relax}%
  \newcommand*\lineheight[1]{\fontsize{\fsize}{#1\fsize}\selectfont}%
  \ifx\svgwidth\undefined%
    \setlength{\unitlength}{340.15748031bp}%
    \ifx\svgscale\undefined%
      \relax%
    \else%
      \setlength{\unitlength}{\unitlength * \real{\svgscale}}%
    \fi%
  \else%
    \setlength{\unitlength}{\svgwidth}%
  \fi%
  \global\let\svgwidth\undefined%
  \global\let\svgscale\undefined%
  \makeatother%
  \begin{picture}(1,0.66666667)%
    \lineheight{1}%
    \setlength\tabcolsep{0pt}%
    \put(0,0){\includegraphics[width=\unitlength,page=1]{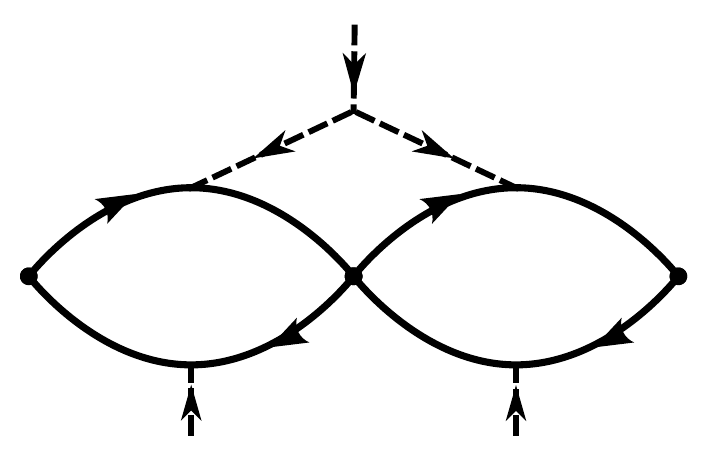}}%
    \put(0.54367486,0.57359851){\color[rgb]{0,0,0}\makebox(0,0)[lt]{\lineheight{1.25}\smash{\begin{tabular}[t]{l}$q_1,\omega_1$\end{tabular}}}}%
    \put(0.30109773,0.08520181){\color[rgb]{0,0,0}\makebox(0,0)[lt]{\lineheight{1.25}\smash{\begin{tabular}[t]{l}$q_2,\omega_2$\end{tabular}}}}%
    \put(0.75943685,0.08520181){\color[rgb]{0,0,0}\makebox(0,0)[lt]{\lineheight{1.25}\smash{\begin{tabular}[t]{l}$q_3,\omega_3$\end{tabular}}}}%
    \put(-0.02645833,0.26939549){\color[rgb]{0,0,0}\makebox(0,0)[lt]{\lineheight{1.25}\smash{\begin{tabular}[t]{l}$k$\end{tabular}}}}%
    \put(0.98998261,0.2696107){\color[rgb]{0,0,0}\makebox(0,0)[lt]{\lineheight{1.25}\smash{\begin{tabular}[t]{l}$k+q_1$\end{tabular}}}}%
    \put(0.54606869,0.2696107){\color[rgb]{0,0,0}\makebox(0,0)[lt]{\lineheight{1.25}\smash{\begin{tabular}[t]{l}$k+q_1+q_3$\end{tabular}}}}%
  \end{picture}%
\endgroup%
 
  }
  \hfill\null
  \caption{Feynman diagrams for
  $\expval{\rho_{(2)}(1)\rho_{(1)}(2)\rho_{(1)}(3)}_{0,c}$.}
  \label{fig:c3diagrams}
\end{figure}

In this section, we calculate the three-point density correlation function using
diagrams of the bosonized theory. For simplicity, we ignore the effects of
interaction. The correlation function is defined as
\begin{equation}
  C_{(3)}(1,2,3)=\expval{\rho(1)\rho(2)\rho(3)}_c,
\end{equation}
where we used the shorthand $i=(q_i,\omega_i)$ and
$\rho(i)=\int_kf(k+q_i,k,\omega_i)$ for $i=1,2,3$ .
We define the $n$-th order term of the expansion of $\rho$ in powers of $\phi$
as $\rho_{(n)}(i)\equiv\int_{k}f_{(n)}(k+q_i,k,\omega_i)$.
In order to expand 
As we showed above,
only the connected tree diagrams need to be considered. Schematically, there is
only one such term as shown in \cref{fig:schm3}, so
\begin{equation}
  C_{(3)}(1,2,3)=\expval{\rho_{(2)}(1)\rho_{(1)}(2)\rho_{(1)}(3)}_{0,c}
  +\expval{\rho_{(1)}(1)\rho_{(2)}(2)\rho_{(1)}(3)}_{0,c}
  +\expval{\rho_{(1)}(1)\rho_{(1)}(2)\rho_{(2)}(3)}_{0,c}.
\end{equation}
We evaluate the first term diagramatically. The other two terms are obtained by
permuting the indices of the arguments. The first term generates two diagrams
that are related by exchange of $(q_2,\omega_2)\leftrightarrow(q_3,\omega_3)$ as
shown in \cref{fig:c3diagrams}. The diagrams evaluate to
\begin{align}
  \expval{\rho_{(2)}(1)\rho_{(1)}(2)\rho_{(1)}(3)}_{0,c}
  =&-\int_k\bigg(\frac{F^{(3)}(k+q_1,k+q_1+q_2,k)}
  {(\omega_2-\varepsilon_{k+q_1+q_2}+\varepsilon_{k+q_1})
  (\omega_3-\varepsilon_{k}+\varepsilon_{k+q_1+q_2})}\notag\\
  &\qquad+\frac{F^{(3)}(k+q_1,k+q_1+q_3,k)}
  {(\omega_2-\varepsilon_k+\varepsilon_{k+q_1+q_3})
  (\omega_3-\varepsilon_{k+q_1+q_3}+\varepsilon_{k+q_1})}\bigg)\notag\\
  &\qquad\times(2\pi)^{d+1}\delta^d(q_1+q_2+q_3)\delta(\omega_1+\omega_2+\omega_3).
  \label{eq:c3a}
\end{align}
The first and second terms correspond to \cref{fig:c3a} and \cref{fig:c3b}
respectively.  By permuting the external indices and reorganizing the terms, we
find
\begin{align}
  C_{(3)}(1,2,3)=&
  -\int_k\bigg(\frac{F^{(3)}(k+q_1,k+q_1+q_2,k)}
  {(\omega_3-\varepsilon_k+\varepsilon_{k+q_1+q_2})
  (\omega_2-\varepsilon_{k+q_1+q_2}+\varepsilon_{k+q_1})}\notag\\
  &\qquad+\frac{F^{(3)}(k+q_2,k+q_2+q_3,k)}
  {(\omega_1-\varepsilon_k+\varepsilon_{k+q_2+q_3})
  (\omega_3-\varepsilon_{k+q_2+q_3}+\varepsilon_{k+q_2})}\notag\\
  &\qquad+\frac{F^{(3)}(k+q_3,k+q_1+q_3,k)}
  {(\omega_2-\varepsilon_k+\varepsilon_{k+q_1+q_3})
  (\omega_1-\varepsilon_{k+q_1+q_3}+\varepsilon_{k+q_3})}+(2\leftrightarrow3)\bigg).
  \label{eq:C3}
\end{align}
Here, we omitted the delta function factors that conserve momentum and
frequency. The first three terms correspond to the fermion loop diagram that
describe the scattering process, $k\rightarrow k+q_1\rightarrow
k+q_1+q_2\rightarrow k$ which we refer to as the $(123)$ scattering process.
The $(2\leftrightarrow3)$ term corresponds to the $(132)$ scattering process
which is the reversed process. In \cref{sec:comparison}, we show that
\cref{eq:C3} exactly matches the corresponding fermion loop calculation.

Notice, that in the low-momentum limit, $F^{(3)}$ in the numerator of each term
vanishes if $k$ is not near the Fermi surface. Moreover, if we take the limit
$\abs{q_i}\rightarrow0$, the $F^{(3)}$ factors becomes a delta function
$\delta(\abs{k}-k_F)$ and the integral over $k$ is reduced to an angular
integral. The leading and sub-leading behavior can be straightforwardly
extracted from this.

Let us study the long-wavelength behavior of the correlation function.  As we
mentioned above, in the original fermionic theory, partial cancellation between
fermion loop diagrams occur when $\omega_i\rightarrow0, q_i\rightarrow0$, and
the long-wavelength behavior of the density
correlation function is determined by the sub-leading term \cite{neumayr1998,
holder2015}.  This cancellation
can be tricky to see in the original theory, but we demonstrate it is
straightforward to see in our bosonic theory. Consider the term \cref{eq:c3a}.
The numerator of the first term expands to
\begin{equation}
  F^{(3)}(k+q_1,k+q_1+q_2,k)=f_{0}(k)-f_0(k+q_2)
  +\qty(\substack{\textrm{terms symmetric under }\\
  \textrm{permutation of }} \qty{q_1,q_1+q_2,q_1+q_2+q_3}).
\end{equation}
The symmetric term vanishes if we consider all other diagrams, so we only need
to consider the first two terms. (This is also explained in
\cref{sec:comparison}.) The leading order behavior of the first term in
\cref{eq:c3a} is
\begin{equation}
  \int_{k}\frac{q_2\cdot\hat k\delta(\abs{k}-k_F)}
  {\qty(\omega_2-q_2\cdot\nabla_k\varepsilon_k)
  \qty(\omega_3+(q_1+q_2)\cdot\nabla_k\varepsilon_k)}.
  \label{eq:c3leadingorder}
\end{equation}
Clearly this diverges as $q_i,\omega_i\rightarrow0$, but notice that the first
and second terms in \cref{eq:c3a} are related by a change in sign of the
external momenta and frequency, $q_i,\omega_i\rightarrow-q_i,-\omega_i$. This
tells us that the leading order behavior of the second term as
$q_{i},\omega_i\rightarrow0$ is given by the negative of
\cref{eq:c3leadingorder}.  Therefore, the leading order behavior of the two
terms in \cref{eq:c3a} cancel and their sum does not diverge in the
long-wavelength limit. Instead, the sub-leading terms of the two terms in
\cref{eq:c3a} determines the scaling. 

\subsection{Four-point density correlation function}
\label{sec:fourpoint}
Let us now consider the four-point density correlation function defined as
\begin{equation}
  C_{(4)}(1,2,3,4)=\expval{\rho(1)\rho(2)\rho(3)\rho(4)}_c.
\end{equation}
The three possible connected tree diagrams are schematically shown in
\crefrange{fig:schm41}{fig:schm43}. They correspond to the terms
\begin{align}
  \begin{split}
  C_{(4)}(1,2,3,4)=&
  \qty(\expval{\rho_{(3)}(1)\rho_{(1)}(2)\rho_{(1)}(3)\rho_{(1)}(4)}_{0,c}+
  \textrm{(other permutations)})\\
  &+ \qty(\expval{\rho_{(2)}(1)\rho_{(2)}(2)\rho_{(1)}(3)\rho_{(1)}(4)}_{0,c}
  +\textrm{(other permutations)})\\
  &+\expval{\rho_{(1)}(1)\rho_{(1)}(2)\rho_{(1)}(3)\rho_{(1)}(4)
  iS_{(4)}}_{0,c}.
  \end{split}
  \label{eq:c4schematic}
\end{align}
To simplify the comparison with the corresponding fermion loop diagram, we
organize the diagrams by the scattering process they describe. There are $3!=6$
distinct scattering processes. For simplicity, we only consider diagrams that
describe the fermion scattering process, $k\rightarrow k+q_1$ $\rightarrow
k+q_1+q_2$ $\rightarrow k+q_1+q_2+q_3$ $\rightarrow k$ which we refer to as the
$(1234)$ scattering process as we did in the previous subsection.  The other
diagrams can be obtained by permuting the external momenta and frequencies. It's
straightforward to iterate through the diagrams that contribute to the $(1234)$
scattering process.

\begin{figure}[h]
  \centering
  \hfill
  \subcaptionbox{\label{fig:c4a}}{
    \def\svgwidth{0.49\textwidth}
    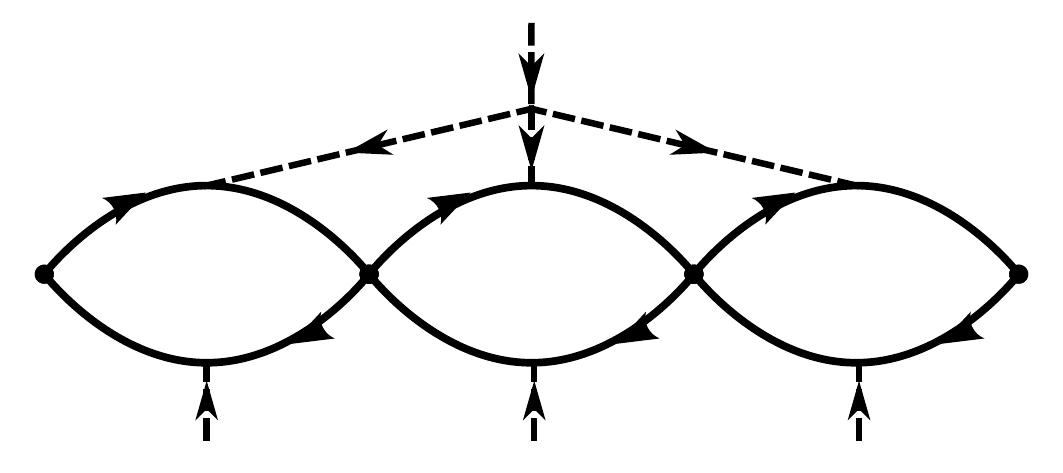 
  }
  \hfill\null

  \bigskip

  \makebox[\textwidth][c]{
  \subcaptionbox{\label{fig:c4b}}{
    \def\svgwidth{0.49\textwidth}
    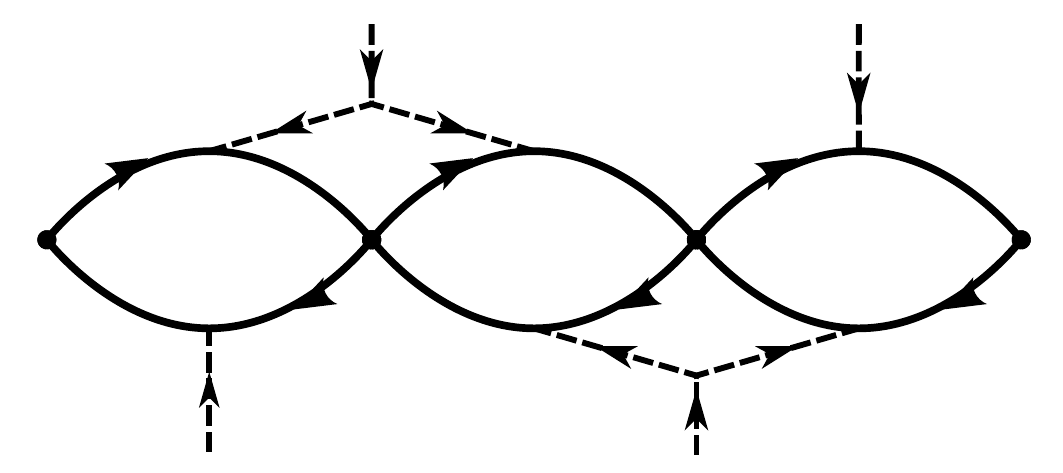 
  }
  \hspace{20pt}
  \subcaptionbox{\label{fig:c4c}}{
    \def\svgwidth{0.49\textwidth}
    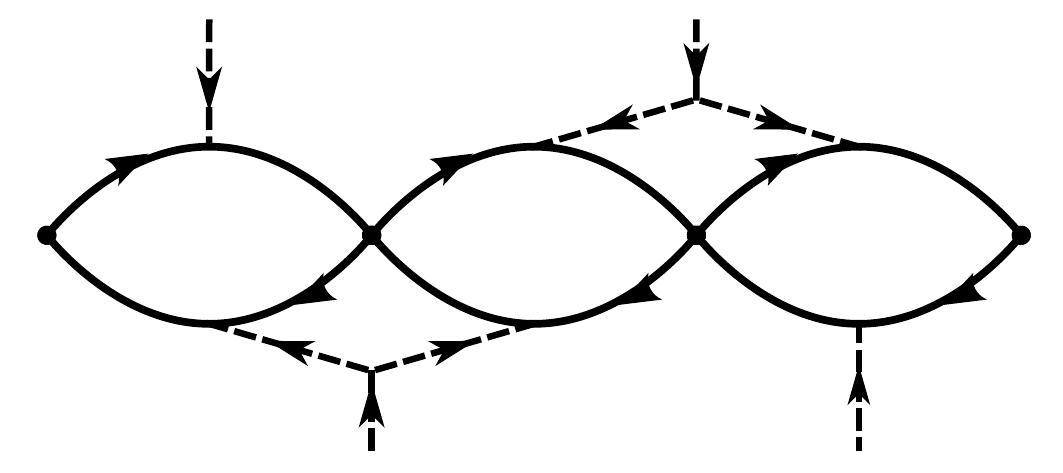 
  }
  }

  \bigskip

  \hfill
  \subcaptionbox{\label{fig:c4d}}{
    \def\svgwidth{0.37\textwidth}
    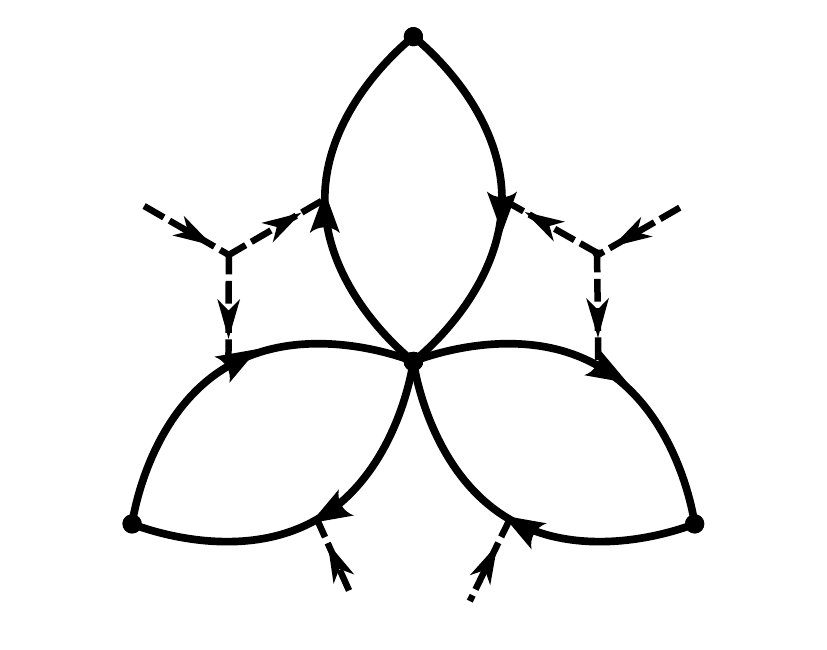 
  }
  \hfill
  \subcaptionbox{\label{fig:c4e}}{
    \def\svgwidth{0.37\textwidth}
    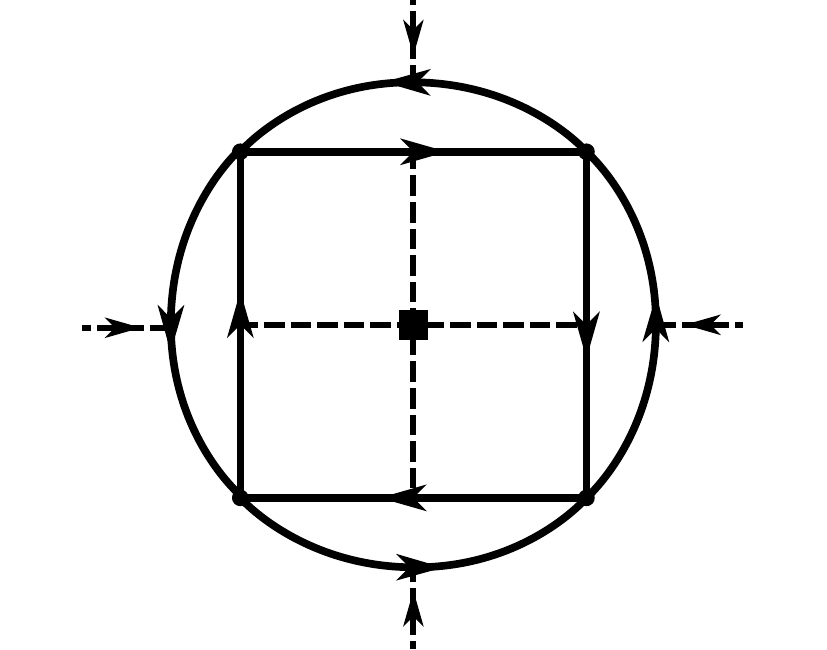 
  }
  \hfill\null

  \caption{Feynman diagrams of $C_{(4)}$ describing the $(1234)$ scattering
  process.  (a) corresponds to terms of the form
  $\expval{\rho_{(3)}\rho_{(1)}\rho_{(1)}\rho_{(1)}}_{0,c}$,
  (b--d) descend from terms of the form
  $\expval{\rho_{(2)}\rho_{(2)}\rho_{(1)}\rho_{(1)}}_{0,c}$, and
  (e) corresponds to
  $\expval{\rho_{(1)}\rho_{(1)}\rho_{(1)}\rho_{(1)}iS_{(4)}}_{0,c}$}
  \label{fig:c4diagrams}
\end{figure}

The first term in \cref{eq:c4schematic} gives us the diagram \cref{fig:c4a}.
The second term leads to three diagrams \crefrange{fig:c4b}{fig:c4c}, and the
last term gives us \cref{fig:c4d}.

Let
\begin{equation}
  g_{ij}(k)=g_{ji}(k)\equiv
  \frac{f_0(k+p_i)-f_0(k+p_j)}
  {\Omega_i-\Omega_j-\varepsilon(k+p_i)+\varepsilon(k+p_j)},
\end{equation}
where $p_i\equiv \sum_{j=1}^iq_j$ and $\Omega_i\equiv\sum_{j=1}^i\omega_i$ for
$i=1,\cdots,4$. The indices are defined mod $4$ for convenience.  All the
diagrams in \cref{fig:c4diagrams} can be expressed as an integral over products
of these functions. Below, we list the contribution from each diagram.
\begin{align}
  \textrm{(\cref{fig:c4a})}
    =&-i\frac{2}{3}\sum_{i=1}^4\int_k
    g_{i,i+1}(k)g_{i+1,i+2}(k)g_{i+2,i+3}(k)\label{eq:c4a}\\
  \textrm{(\cref{fig:c4b})}
    =&-i\sum_{i=1}^2\int_k g_{i+3,i+2}(k)g_{i+2,i}(k)g_{i,i+1}(k)\\
  \textrm{(\cref{fig:c4c})}
    =&-i\sum_{i=1}^2\int_k g_{i,i+1}(k)g_{i+1,i+3}(k)g_{i+3,i+2}(k)\\  
  \textrm{(\cref{fig:c4d})}
    =&i\sum_{i=1}^4\int_k g_{i+1,i}(k)g_{i+2,i}(k)g_{i+3,i}(k)\\
  \textrm{(\cref{fig:c4e})}
    =&\frac{i}{6}\sum_{i=1}^4\int_k
    g_{i,i+1}(k)g_{i+1,i+2}(k)g_{i+2,i+3}(k)\label{eq:c4e}
\end{align}
The factor $(2\pi)^{d+1}\delta^d(q_1+q_2+q_3+q_4)
\delta(\omega_1+\omega_2+\omega_3+\omega_4)$ was omitted. Notice, that
\cref{eq:c4a} and
\cref{eq:c4e} are identical up to an overall factor. 
Adding all these contributions together gives us
\begin{align}
  C_{(4)}(1,2,3,4)=&\bigg(-\frac{i}{2}\sum_{i=1}^4\int_k
  g_{i+3,i+2}(k)g_{i+2,i+1}(k)g_{i+1,i}(k)
 +i\sum_{i=1}^4\int_k
  g_{i+1,i}(k)g_{i+2,i}(k)g_{i+3,i}(k)\notag\\
  &-i\sum_{i=1}^2\int_k g_{i+3,i+2}(k)g_{i+2,i}(k)g_{i,i+1}(k)
    -i\sum_{i=1}^2\int_k g_{i+2,i+3}(k)g_{i+3,i+1}(k)g_{i+1,i}(k)\bigg)\notag\\
    &+(\textrm{permutations of }
    \{(\omega_1,q_1),(\omega_2,q_2),(\omega_3,q_3),
    (\omega_4,q_4)\})
    \label{eq:C4}
\end{align}
where we omitted the overall delta function factors that conserve momentum and
frequency.  The first four terms correspond to the $(1234)$ scattering process.
In \cref{sec:comparison}, we show that \cref{eq:C4} equals the corresponding
fermion loop calculation exactly.

\subsection{Self-energy calculations}
Up till now we have ignored interaction, so let us now consider its effects.  As
we discussed above, a general two-particle interaction introduces the term
\cref{eq:interaction} to the action.  Expanding in powers of $\phi$, we define
\begin{equation}
  S_{\textrm{int}}^{(n,n')}=S_{\textrm{int}}^{(n',n)}
  =-\frac{1}{2}\int_{q,k,k',t}
  \tilde{V}_qf_{(n)}(k+q,k,t)f_{(n')}(k'-q,k',t)
\end{equation}
so that $S_{\textrm{int}}=\sum_{n,n'=1}^\infty S_{\textrm{int}}^{(n,n')}$.
Let us now think about what kind of diagrams can be drawn if we include
these interaction terms. The interaction term is introduced by taking a
second-order functional derivative with respect to $J$, so diagrammatically,
interactions can be thought of as a diagram element that stitches together two
external legs of a tree diagram. Here, when we refer to external legs we refer
to the dashed lines in the full diagrams or the points in the schematic
diagrams.  The external legs can be from the same tree diagram or from different
tree diagrams.

\begin{figure}[h]
  \centering

  \hfill
  \subcaptionbox{}{
    \def\svgwidth{0.4\textwidth}
    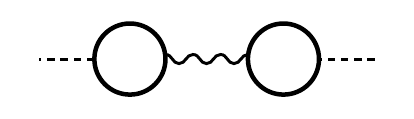
  }
  \hfill
  \subcaptionbox{}{
    \def\svgwidth{0.4\textwidth}
    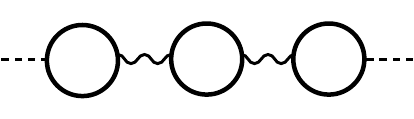
  }
  \hfill\null\\

  \hfill
  \subcaptionbox{}{
    \def\svgwidth{0.4\textwidth}
    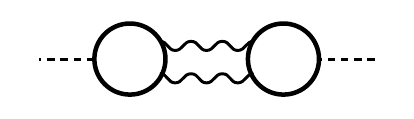
  }
  \hfill
  \subcaptionbox{}{
    \def\svgwidth{0.4\textwidth}
    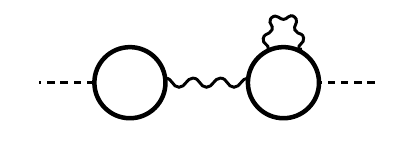
  }
  \hfill\null

  \caption{Schematic representations of the corrections to the two-point density
  correlation function up to second order in interaction. Here, we used a
  different diagrammatic convention from before. Circles represent the bare
  two-, three-, and four-point density correlation functions depending on the
  number of legs they have. The wavy lines represent the stitching of two legs
  by interaction.}
  \label{fig:intcncorrection}
\end{figure}

For simplicity, let us consider the two-point density correlation function. The
corrections to the correlation function up to second order in interaction are
shown in \cref{fig:intcncorrection}. Each circle, depending on the number of
legs it possesses, represents a two-, three-, or four-point correlation function
of the \emph{bare} theory. It is clear, that higher-order corrections will
involve higher-order correlation functions of the bare fermion.

Let us consider a simple approximation where we only consider diagrams that
involve two-point correlation functions. It is then straightforward to show that
the two-point density correlation function of the interacting theory under this
approximation is
\begin{align}
  C_{(2),\textrm{int}}(1,2)&\approx \int_kD_0(k+q_1,k,\omega_1)
  \sum_{n=0}^\infty
  \qty(-i\tilde V_q\int_kD_0(k+q_1,k,\omega_1))^{n}\\
  &=\frac{\int_k D_0(k+q_1,k,\omega_1)}
  {1+i\tilde V_q\int_k D_0(k+q_1,k,\omega_1)}
\end{align}
where the momentum and frequency delta functions are omitted. This is equivalent
to the RPA approximation. 

\section{Discussion}
\label{sec:discussion}
In this work, we derived an exact method to express a fermionic theory with a
Fermi surface as a restricted functional integral over a bosonic field. Starting
from a generating functional with a source coupled to density, we calculated the
Legendre transform of the free energy. We then showed that the integral of the
Legendre-transformed free energy over the Lie algebra of the unitary group when
restricted to tree level diagrams reproduces the generating functional. In order
to demonstrate the utility of our formalism, we evaluated the three- and
four-point density correlation functions using Feynman diagram rules that we
proposed. The results from our bosonized theory exactly matches results obtained
from fermion loop calculations.

This work was inspired by the work by Delacretaz et. al.  \cite{delacretaz2022}.
Using heuristic arguments, they derived an elegant bosonization scheme for a
system with a Fermi surface that incorporates nonlinear effects. It turns out
that the long-wavelength limit of the bosonized action that we derived,
\cref{eq:action}, is equivalent to the bosonic action in Ref.
\cite{delacretaz2022}.  We can see this correspondence by expressing our
bosonized action in the Wigner representation.  In this representation, operator
products are replaced by the non-commutative Moyal product. The action derived
by Delacretaz et. al. can then be obtained by expanding the Moyal product to
first order in spatial derivatives.  Then, the Lie algebra of the unitary group
becomes the Poisson algebra of the canonical transformations and the two actions
become equivalent. Therefore, our work serves as a rigorous derivation of the
results presented in Ref.  \cite{delacretaz2022}. In \cref{sec:correspondence}
we go over this argument in more detail.

In addition to providing a proof, our work also clarifies ambiguities in Ref.
\cite{delacretaz2022} that we listed earlier in \cref{sec:intro}.  First of all,
the bosonized theory in Ref. \cite{delacretaz2022} is valid in the
long-wavelength limit, so a question one can ask is how can we obtain
corrections to this limit?  As discussed above, the long-wavelength limit is a
result of using the Poisson algebra so corrections can be obtained by expanding
to the Lie algebra of the unitary group. Our bosonized action is exact and
considers fluctuations of the Fermi surface generated by the unitary group, so
our action can be be used to obtain these corrections. Secondly, Ref.
\cite{delacretaz2022} does not explain why correlation functions can be
calculated from only tree diagrams. We explained in \cref{sec:derivation} that
this is due to taking the saddle-point approximation to obtain the Legendre
transform.  Lastly, Ref. \cite{delacretaz2022} did not explicitly derive the
link between a general four-fermion interaction term and the interaction term in
their bosonized action. We derived this in \cref{sec:derivation}.

The bosonization formalism in this paper adapts an old field theory result
concerning the effective action typically found in field theory textbooks
\cite{weinberg1995}. Naturally, one might ask what is new in this work. This
result on the effective action usually applies to bosonic systems with a source
coupled linearly to the field, and the Legendre transformation of this action
produces another bosonic action.  However, in this work we started with a
fermionic theory and coupled the source to a fermion bilinear so that the
Legendre transform we calculated produces a bosonic action from a fermionic
action.  In addition, the change of variables from the density matrices to the
Lie algebra of the unitary group is a non-linear transformation that partitions
the original fermion diagrams in a non-trivial manner.  These aspects of our
formalism are well beyond the scope of the aforementioned textbook result.

In many ways the bosonization formulation derived in this paper appears similar
to standard one-dimensional (1D) bosonization. For one thing, if we take the
second-order contribution to the action in 1D and express it in the Wigner
representation, to lowest order of the gradient expansion we get
$S_{(2)}\approx\frac{1}{2}\sum_{s=\pm}\int_{t,x}(\partial_x\phi_s\partial_t\phi_s
+v_F(\partial_x\phi_s)^2)$ where $\phi_\pm(x,t)\equiv\phi(x,\pm k_F, t)$
\cite{delacretaz2022}. This is the bosonized action of non-interacting 1D
fermions with linear dispersion \cite{haldane1981}. In addition, the
calculation of an equal-time Green's function in our formalism, schematically,
is $G\sim\expval{\psi^\dagger\psi}
\sim\expval{f}\sim\expval{e^{i\phi}f_0e^{-i\phi}}$ for any dimension.
The last expression is
reminiscent of a Green's function calculation in standard 1D bosonization where
$e^{\pm i \phi}$ plays the role of a vertex operator. However, despite their
similarities, the two bosonization formalisms are distinct.  Under standard 1D
bosonization, the action of 1D fermions with an exactly linear dispersion is
strictly quadratic, but under our bosonization formalism, it will generally have
an infinite number of higher-order terms.  In addition, our formalism has an
artificially introduced parameter $\alpha$ that is absent in standard 1D
bosonization.  This means that in our formalism
$G\sim\expval{e^{i\phi}f_0e^{-i\phi}}$ cannot be naively evaluated as an
exponential of the propagator of $\phi$ and the anomalous dimension cannot be
extracted this way as is done in 1D bosonization.  Therefore, it is clear that
the two approaches are distinct despite some similarities, and further
investigation is required to better understand their connection.

In our work, we briefly discussed the effects of interaction, and we obtained
the RPA approximation by only considering the simplest series of diagrams that
contribute to the two-point density correlation function. In general, because
the bosonization scheme partitions fermion diagrams into several bosonic
diagrams, this increases the variety of diagrams that we need to consider in the
bosonized theory. In future works, it would be of value to consider all of
these diagrams and see how they can be organized differently compared to the
original fermionic theory. 

Another direction to consider in the future is to apply the bosonization method
we outlined in this work to other systems with different relevant degrees of
freedom.  In principle, our approach can be simply generalized to different
cases by coupling the source $J$ in the generating functional to different
fermion bilinears. For example, $J$ can couple to  $\psi(x,t)\psi(x',t)$ to
describe a BCS theory or to $\psi_\alpha^{\dagger}(x,t)
\frac{\vec{\sigma}_{\alpha\beta}}{2} \psi_{\beta}(x',t)$ where
$\alpha,\beta=\uparrow,\downarrow$ to describe spin fluctuations. One can also
consider the case where the source is coupled to two fermion operators at
different times, $\psi^\dagger(x,t)\psi(x',t')$. This was considered by Han et.
al. in their proposal of a bosonization procedure for non-Fermi liquids in Ref.
\cite{han2023}.

\section*{Acknowledgements}
The authors thank Yi-Hsien Du, L\'{e}o Mangeolle, and Lucile Savary for
insightful discussions.
T.P. was supported by a quantum Foundry fellowship through the National Science Foundation through Enabling
Quantum Leap: Convergent Accelerated Discovery Foundries for Quantum Materials
Science, Engineering and Information (Q-AMASE-i) award number
DMR-1906325, supplemented by the  NSF CMMT
program under Grant No. DMR-2116515, which also supported  L.B.

\begin{appendix}
\section{Obtaining the action derived by Delacretaz et.  al. \cite{delacretaz2022}
from \cref{eq:action}}
\label{sec:correspondence}
Here, we show how to explicitly obtain the action derived by Delacretez et.
al.in Ref. \cite{delacretaz2022}. For convenience we replicate the action in
\cref{eq:action}
here.
\begin{equation}
  S[\phi]\equiv
  -\Gamma[f[\phi]]+\Gamma[f_0]
  =\Tr[f_0e^{-i\phi}(i\partial_t-H_0)e^{i\phi}]+\Tr[f_0H_0].
\end{equation}

To obtain the action in Ref. \cite{delacretaz2022}, we need to express the
equation above in the Wigner representation. The Wigner representation is a
mixed basis representation.  Given a single-particle operator $\hat A$, its Wigner
representation is defined as \cite{zachos2005, curtright2014}
\begin{equation}
  \hat A\rightarrow A(x,p)=\int_{y}e^{-iy\cdot p}
  \mel{x+\frac{y}{2}}{\hat A}{x-\frac{y}{2}}.
\end{equation}
Here, a hat is added to operators to emphasize their distinction from c-numbers,
but in the main text we do not.  The product of operators can be represented as
the Moyal product ($\star$-product) of their Wigner representations.
\begin{equation}
  \hat A\hat B\rightarrow A(x,p)\star B(x,p)\equiv
  A(x,p)e^{\frac{i}{2}\qty(\cev{\nabla}_x\cdot\vec{\nabla}_p-
  \cev{\nabla}_p\cdot\vec{\nabla}_x)}B(x,p)
\end{equation}
The arrows of the derivatives indicate the directions in which they act.

The trace of an operator is the integral of its Wigner representation over phase
space.
\begin{equation}
  \tr[\hat A]=\int_{x,p}A(x,p)
\end{equation}
The trace of the product of two operators is the phase-space integral of the
products of their Wigner representation.
\begin{equation}
  \tr[\hat A\hat B]=\int_{x,p}A(x,p)B(x,p)
  \label{eq:traceproperty}
\end{equation}

The commutator of two operators in the Wigner representation becomes the Moyal
bracket.
\begin{equation}
  \comm{\hat A}{\hat B}\rightarrow \mpb{A(x,p)}{B(x,p)}=A(x,p)\star B(x,p)-B(x,p)\star
  A(x,p).
\end{equation}
The Moyal bracket defines a Lie algebra which we refer to as the
$\star$-algebra.  If $A(x,p)$ and $B(x,p)$ vary slowly in space, we can expand
the Moyal product in powers of the gradient. The lowest order term of the
commutator in the gradient expansion turns out to be the Poisson bracket.
\begin{equation}
  \mpb{A(x,p)}{B(x,p)}\approx A(x,p)i\qty(\cev{\nabla}_x\cdot\vec{\nabla}_p-
  \cev{\nabla}_p\cdot\vec{\nabla}_x)
  B(x,p)=i\pb{A(x,p)}{B(x,p)}_{\textrm{p.b.}}
\end{equation}
Hence, the Poisson algebra is obtained from the $\star$-algebra in the
long-wavelength limit.

Now that we have reviewed the necessary facts on the Wigner representation needed
to proceed, let us consider the Hamiltonian term of the action. As we discussed
in \cref{sec:diagrams}, it can be expanded using the adjoint action of
$\phi$.
\begin{equation}
  e^{-i\phi}H_0e^{i\phi}=\sum_{n=0}\frac{(-i)^n}{n!}\textrm{ad}^n_{\phi}{H_0}.
\end{equation}
In the Wigner representation this becomes
\begin{equation}
  (e^{-i\phi}H_0e^{i\phi})(x,p)
  =\sum_{n=0}^{\infty}\frac{(-i)^n}{n!}\qty(\textrm{ad}^{(\star)}_{\phi(x,p)})^n{H_0(p)}
\end{equation}
where $\textrm{ad}^{(\star)}_{\phi(x,p)}(\cdot)\equiv \mpb{\phi(x,p)}{\cdot}$ is the
adjoint action of the $\star$-algebra. Assuming $\phi(x,p)$ varies slowly in
space, this becomes
\begin{equation}
  (e^{-i\phi}h_0e^{i\phi})(x,p)\rightarrow
  \sum_{n=0}^{\infty}\frac{1}{n!}\qty(\textrm{ad}^{(p)}_{\phi(x,p)})^n{h_0(p)}
\end{equation}
where $\textrm{ad}^{(p)}_{\phi(x,p)}(\cdot)\equiv
\pb{\phi(x,p)}{\cdot}_{\textrm{p.b.}}$. Following the notation $U=e^{-\phi}$
used in ref. \cite{delacretaz2022}, we get
\begin{equation}
  (e^{-i\phi}H_0e^{i\phi})(x,p)\rightarrow U^{-1}H_0 U.
\end{equation}
where the right-hand side is to be interpreted as the adjoint action of the lie
group that represents the canonical transformations. The same argument follows
for the term with the time-derivative, so
\begin{equation}
  (e^{-i\phi}(i\partial_t-H_0)e^{i\phi})(x,p)\rightarrow U^{-1}(\partial_t-H_0)U.
\end{equation}
using the notation in Ref. \cite{delacretaz2022}, that the integral over the
phase-space can be expressed as an inner-product and using
\cref{eq:traceproperty}, the action becomes
\begin{equation}
  S[\phi]=\tr[f_0e^{-i\phi}(i\partial_t-H_0)e^{i\phi}]\rightarrow 
  \int\dd t\left< f_0,  U^{-1}(\partial_t-H_0)U\right>
\end{equation}

Note, this is not an equality since we assumed that $\phi(x,p)$ varies slowly in
space. This assumption is equivalent to the fact that in the momentum
representation, the field $\phi(k,k',t)$ is non-negligible only when
$\abs{k-k'}\ll k_f$.  When calculating the correlation functions, it can also be
understood as the small momentum limit, $\abs{q_i}\rightarrow0$.

\section{Comparison of the correlation functions obtained from fermion loop
diagrams and the bosonized theory}
\label{sec:comparison}
Here, we show that the three-point and four-point density correlation functions
obtained from our bosonized theory matches the results from fermion loop
calculations.  First, we review the calculation of a fermion loop with $n$
external boson legs.

We begin with the $n$-point density correlation function defined as
\begin{equation}
  C_{(n)}(q_1,\omega_1;\cdots;q_n,\omega_n)
  \equiv \expval{\mathbb{T}\rho(q_1,\omega_1)\cdots\rho(q_n,\omega_2)}_c.
\end{equation}
The density operator is defined as
$\rho(q,\omega)=\int_{k,\omega'}\psi^\dagger(k+q,\omega'+\omega)\psi(k,\omega')$.
$C_{(n)}$ can be expressed as the sum over $(n-1)!$ unique fermion loops. For
simplicity, we only consider the fermion loop that describes the scattering
process $k\rightarrow k+q_1\rightarrow k+q_1+q_2\rightarrow\cdots\rightarrow
k+q_1+\cdots+q_n\rightarrow k$ and denote its contribution as $\mathscr{F}_{(n)}$.

For convenience, we define $p_i\equiv q_1+\cdots+q_i$ and $\Omega_i\equiv
\omega_1+\cdots+\omega_i$. Then, 
\begin{equation}
  \mathscr{F}_{(n)}=-i^n\int_{k,\omega} \prod_{i=1}^nG(k+p_i,\omega+\Omega_i).
\end{equation}
The Green's function is defined as
$G(k,\omega)=-i\expval{T\psi(k,\omega)\psi^\dagger(k,\omega)}
=\qty(\omega-\varepsilon_k+i\textrm{sgn} \varepsilon_k)^{-1}$.
Integrating over $\omega$ we get
\begin{equation}
  \mathscr{F}_{(n)}=-i^{n+1}\int_k\sum_{i=1}^nf_0(k+p_i)\prod_{\substack{j=1\\j\neq i}}^n
  \frac{1}{\Omega_j-\Omega_i-\varepsilon(k+p_j)+\varepsilon(k+p_i)}.
  \label{eq:fermionloopeq}
\end{equation}

\subsection{Three-point correlation function}
Now, let us compare the $n=3$ fermion loop result with the three-point correlation
function obtained in \cref{sec:threepoint}. For convenience, we reproduce the
result given by \cref{eq:C3} below.
\begin{align}
  C_{(3)}(1,2,3)=&
  -\int_k\bigg(\frac{F^{(3)}(k+q_1,k+q_1+q_2,k)}
  {(\omega_3-\varepsilon_k+\varepsilon_{k+q_1+q_2})
  (\omega_2-\varepsilon_{k+q_1+q_2}+\varepsilon_{k+q_1})}\notag\\
  &\qquad+\frac{F^{(3)}(k+q_2,k+q_2+q_3,k)}
  {(\omega_1-\varepsilon_k+\varepsilon_{k+q_2+q_3})
  (\omega_3-\varepsilon_{k+q_2+q_3}+\varepsilon_{k+q_2})}\notag\\
  &\qquad+\frac{F^{(3)}(k+q_3,k+q_1+q_3,k)}
  {(\omega_2-\varepsilon_k+\varepsilon_{k+q_1+q_3})
  (\omega_1-\varepsilon_{k+q_1+q_3}+\varepsilon_{k+q_3})}+(2\leftrightarrow3)\bigg)
\end{align}
We claim that the first three terms, which we label $\mathscr{B}_{(3)}$,
corresponds to the fermion loop $\mathscr{F}_{(3)}$. We prove this by
explicitly showing $\mathscr{B}_{(3)}=\mathscr{F}_{(3)}$.

First, let us express $\mathscr{B}_{(3)}$ using $p_i$ and $\Omega_i$. To do this,
we need to shift the integration variables of the second and third terms by
$k\rightarrow k+q_1$ and $k\rightarrow k+q_1+q_2$ respectively. Then, we use the
fact that $q_i=p_i-p_{i-1}$ and $\omega_i=\Omega_i-\Omega_{i-1}$, where here the
indices are defined mod 3.  This gives us
\begin{equation}
  \begin{split}
  \mathscr{B}_{(3)}=&\int_k\bigg(\frac{F^{(3)}(k+p_1,k+p_2,k+p_3)}
  {(\Omega_3-\Omega_2-\varepsilon_{k+p_3}+\varepsilon_{k+p_2})
  (\Omega_1-\Omega_2-\varepsilon_{k+p_1}+\varepsilon_{k+p_2})}\\
  &\qquad+\frac{F^{(3)}(k+p_2,k+p_3,k+p_1)}
  {(\Omega_1-\Omega_3-\varepsilon_{k+p_1}+\varepsilon_{k+p_3})
  (\Omega_1-\Omega_3-\varepsilon_{k+p_1}+\varepsilon_{k+p_3})}\\
  &\qquad+\frac{F^{(3)}(k+p_3,k+p_1,k+p_2)}
  {(\Omega_2-\Omega_1-\varepsilon_{k+p_2}+\varepsilon_{k+p_1})
  (\Omega_3-\Omega_1-\varepsilon_{k+p_3}+\varepsilon_{k+p_1})}
\end{split}
\label{eq:b3step1}
\end{equation}
Next, let us expand the expression for the $F^{(3)}$ factors. In general,
\begin{equation}
  F^{(3)}(k_a,k_b,k_c)=-f_0(k_b)+f_0(k_a)f_0(k_b)+f_0(k_c)f_0(k_a)+f_0(k_b)f_0(k_c)
  -2f_0(k_a)f_0(k_b)f_0(k_c).
\end{equation}
If we substitute this into \cref{eq:b3step1}, we get
\begin{equation}
  \begin{split}
    \mathscr{B}_{(3)}=&\int_k\bigg[-\sum_{i=1}^3 f_0(k+p_i)
    \prod_{\substack{j=1\\j\neq i}}^3
    \frac{1}{\Omega_j-\Omega_i-\varepsilon_{k+p_j}+\varepsilon_{k+p_i}}\\
    &\qquad+\bigg(f_0(k+p_1)f_0(k+p_2)+f_0(k+p_2)f_0(k+p_3)+f_0(k+p_3)f_0(k+p_1)\\
    &\qquad -2f_0(k+p_1)f_0(k+p_2)f_0(k+p_3)\bigg)
    \underbrace{
    \sum_{i=1}^3 \prod_{\substack{j=1\\j\neq i}}^3\frac{1}
    {\Omega_j-\Omega_i-\varepsilon_{k+p_j}+\varepsilon_{k+p_i}}}_{=0}
    \bigg].
  \end{split}
\end{equation}
It is straightforward to verify that the last factor 
vanishes by explicitly writing it out. It is a sum of products of fractions, and
if we rewrite it as a single fraction with a common denominator, the numerator
will be zero. Therefore, we are left with only the first term in the square
brackets. A quick comparison with \cref{eq:fermionloopeq} shows us that this is
the expression for $\mathscr{F}_{(3)}$:
\begin{equation}
  \mathscr{B}_{(3)}=\mathscr{F}_{(3)}=-\int_k\sum_{i=1}^3 f_0(k+p_i)
    \prod_{\substack{j=1\\j\neq i}}^3
    \frac{1}{\Omega_j-\Omega_i-\varepsilon_{k+p_j}+\varepsilon_{k+p_i}}.
\end{equation}

\subsection{Four-point correlation function}
Let us now check that the four-point correlation function obtained in
\cref{sec:fourpoint} matches the $n=4$ fermion loop result. For convenience, we
reproduce the result given by \cref{eq:C4} below.
\begin{align}
  C_{(4)}(1,2,3,4)=&\bigg(
  \underbrace{-\frac{i}{2}\sum_{i=1}^4\int_k
  g_{i+3,i+2}(k)g_{i+2,i+1}(k)g_{i+1,i}(k)}
  _{\displaystyle\equiv\mathscr{B}_{(4),1}}
 +\underbrace{i\sum_{i=1}^4\int_k
 g_{i+1,i}(k)g_{i+2,i}(k)g_{i+3,i}(k)}
 _{\displaystyle\equiv\mathscr{B}_{(4),2}}\notag\\
 &\underbrace{-i\sum_{i=1}^2\int_k g_{i+3,i+2}(k)g_{i+2,i}(k)g_{i,i+1}(k)}
 _{\displaystyle\equiv\mathscr{B}_{(4),3}}\quad
    \underbrace{-i\sum_{i=1}^2\int_k g_{i+2,i+3}(k)
    g_{i+3,i+1}(k)g_{i+1,i}(k)}
    _{\displaystyle\equiv\mathscr{B}_{(4),4}}\bigg)\notag\\
    &+(\textrm{permutations of }
    \{(\omega_1,q_1),(\omega_2,q_2),(\omega_3,q_3),
    (\omega_4,q_4)\})
    \label{eq:C4copy}
\end{align}
where
\begin{equation}
  g_{i,j}(k)=\frac{f_0(k+p_i)-f_0(k+p_j)}
  {\Omega_i-\Omega_j+\varepsilon_{k+p_i}-\varepsilon_{k+p_j}}.
\end{equation}
Since we are only interested in terms that we want to compare with
$\mathscr{F}_{(4)}$, we can ignore the permuted terms and only consider the
contributions from the first four terms of \cref{eq:C4copy} which we label
$\mathscr{B}_{(4)}$.  The four terms have been separately labelled
$\mathscr{B}_{(4),i}$ for $i=1,2,3,4$. Below, we show that
$\mathscr{B}_{(4)}=\sum_{i=1}^4\mathscr{B}_{(4),i}=\mathscr{F}_{(4)}$.

Let us first look at $\mathscr{B}_{(4),2}$. The denominator of its integrands can
be rewritten as
\begin{align}
  &\bigg(\textrm{Denominator of } g_{i+1,i}(k)g_{i+2,i}(k)g_{i+3,i}(k)\bigg)\\
  &\quad=
    (\Omega_{i+1}-\Omega_i+\varepsilon_{k+p_{i+1}}-\varepsilon_{k+p_i})
    (\Omega_{i+2}-\Omega_i+\varepsilon_{k+p_{i+2}}-\varepsilon_{k+p_i})
    (\Omega_{i+3}-\Omega_i+\varepsilon_{k+p_{i+3}}-\varepsilon_{k+p_i})
    \notag\\
    &\quad=\prod_{\substack{j=1\\j\neq i}}^4
    \qty(\Omega_j-\Omega_i+\varepsilon_{k+p_j}-\varepsilon_{k+p_i}).
\end{align}
This is the denominator in $\mathscr{F}_{(4)}$.  Let us now consider the
numerator of the integrands in $\mathscr{B}_{(4),2}$.
\begin{equation}
  \begin{split}
    &\bigg(\textrm{Numerator of } g_{i+1,i}(k)g_{i+2,i}(k)g_{i+3,i}(k)\bigg)\\
  &\qquad=(f_0(k+p_{i+1})-f_0(k+p_i))(f_0(k+p_{i+2})-f_0(k+p_i))
  (f_0(k+p_{i+3})-f_0(k+p_i))\\
  &\qquad=-f_0(k+p_i)
  +f_0(k+p_i)\bigg[f_0(k+p_{i+1})+f_0(k+p_{i+2})+f_0(k+p_{i+3})\\
  &\qquad\quad -f_0(k+p_{i+1})f_0(k+p_{i+2})-f_0(k+p_{i+1})f_0(k+p_{i+3})
  -f_0(k+p_{i+2})f_0(k+p_{i+3})\bigg]
  \end{split}
  \label{eq:numexpansion}
\end{equation}
If we separate the first term of the above expansion from the rest,
$\mathscr{B}_{(4),2}$ becomes
\begin{align}
    \mathscr{B}_{(4),2}=&i\sum_{i=1}^4\int_k
    \qty(-f_0(k+p_i)+\textrm{(other terms)})\prod_{\substack{j=1\\j\neq i}}^4
    \frac{1}{\Omega_j-\Omega_i+\varepsilon_{k+p_j}-\varepsilon_{k+p_i}}\\
    =&\mathscr{F}_{(4)}+\underbrace{i\sum_{i=1}^4\int_k
    \textrm{(other terms)}\prod_{\substack{j=1\\j\neq i}}^4
    \frac{1}{\Omega_j-\Omega_i+\varepsilon_{k+p_j}-\varepsilon_{k+p_i}}}
    _{\displaystyle\equiv \mathcal{B}_{(4),2,R}}.
\end{align}
Therefore, the first term of the expansion given in
\cref{eq:numexpansion} gives us $\mathscr{F}_{(4)}$. The remaining terms in
$\mathscr{B}_{(4),2}$ are collectively labelled $\mathscr{B}_{(4),2,R}$.
Although tedious, it can then be shown by taking the integrands of 
$\mathscr{B}_{(4),2,R}, \mathscr{B}_{(4),1}, \mathscr{B}_{(4),3},
\mathscr{B}_{(4),4}$ and rewriting them as a single fraction with a common
denominator that
\begin{equation}
  \mathscr{B}_{(4),2,R}+\mathscr{B}_{(4),1}+\mathscr{B}_{(4),3}+\mathscr{B}_{(4),4}=0.
\end{equation}
Therefore,
\begin{equation}
  \mathscr{B}_{(4)}=
  \mathscr{B}_{(4),1}+\mathscr{B}_{(4),2}+\mathscr{B}_{(4),3}+\mathscr{B}_{(4),4}
  =\mathscr{F}_{(4)}.
\end{equation}

\end{appendix}


\bibliography{references.bib}

\nolinenumbers

\end{document}